\documentclass[twocolumn,showpacs,preprintnumbers,amsmath,amssymb]{revtex4}
\usepackage{epsfig}
\usepackage{dcolumn}
\usepackage{bm}

\newcommand{\beq}{\begin{equation}} 
\newcommand{\eeq}{\end{equation}} 
\newcommand{\beqa}{\begin{eqnarray}} 
\newcommand{\eeqa}{\end{eqnarray}} 
\newcommand{\bea}{\begin{array}} 
\newcommand{\ea}{\end{array}} 

\newcommand{\dd}{{\rm d}}
\renewcommand{\pl}{\partial}
 
\newcommand{\lag}{\langle} 
\newcommand{\rag}{\rangle}

\newcommand{\ii}{{\rm i}}

\newcommand{\ve}{{\bf e}}
\newcommand{\vn}{{\bf n}}
\newcommand{\vu}{{\bf u}}
\newcommand{\vv}{{\bf v}}

\newcommand{\vx}{{\bf x}}
\newcommand{\vk}{{\bf k}}
\newcommand{\vq}{{\bf q}}
\newcommand{\vs}{{\bf s}}
\newcommand{\vQ}{{\bf Q}}
\newcommand{\vX}{{\bf X}}
\newcommand{\vU}{{\bf U}}
\newcommand{\vK}{{\bf K}}

\newcommand{\tpsi}{{\tilde{\psi}}}
\newcommand{\tdelta}{{\tilde{\delta}}}
\newcommand{\tkappa}{{\tilde{\kappa}}}

\newcommand{\pR}{{G}}

\newcommand{\tR}{{\tilde{G}}}
\newcommand{\tG}{{\tilde{G}}}

\newcommand{\tS}{{\tilde{S}}}

\newcommand{\tW}{{\tilde{W}}}
\newcommand{\mA}{{\cal A}}

\newcommand{\cD}{{\cal D}}

\newcommand{\cL}{{\cal L}}
\newcommand{\cF}{{\cal F}}
\newcommand{\cO}{{\cal O}}

\newcommand{\cV}{{\cal V}}

\newcommand{\lvec}[1]{\overrightarrow{#1}}

\begin{document}


\title{Exact Results for Propagators in the Geometrical Adhesion Model}

\author{Francis Bernardeau}
\affiliation{Institut de Physique Th\'eorique,\\
CEA, IPhT, F-91191 Gif-sur-Yvette, C\'edex, France\\
CNRS, URA 2306, F-91191 Gif-sur-Yvette, C\'edex, France}
\author{Patrick Valageas}
\affiliation{Institut de Physique Th\'eorique,\\
CEA, IPhT, F-91191 Gif-sur-Yvette, C\'edex, France\\
CNRS, URA 2306, F-91191 Gif-sur-Yvette, C\'edex, France}
\vspace{.2 cm}

\date{\today}
\vspace{.2 cm}

\begin{abstract}
The Geometrical Adhesion Model (GAM) that we described in previous papers provides a 
fully solved model for the nonlinear evolution of fields that mimics the cosmological 
evolution of pressureless fluids. In this context we explore the expected late time 
properties of the cosmic propagators once halos have formed, in a regime beyond the 
domain of application of perturbation theories. Whereas propagators in Eulerian 
coordinates are closely related to the velocity field we show here that propagators defined 
in Lagrangian coordinates are intimately related to the halo mass function. Exact results
can be obtained in the 1D case. In higher dimensions, the computations are more 
intricate because of the dependence of the propagators on
the detailed shape
of the underlying Lagrangian-space tessellations, that is, on the geometry of the
regions that eventually collapse to form halos.
We illustrate these results for both the 1D and the 2D dynamics. In particular we give 
here the expected asymptotic behaviors obtained for power-law initial power spectra. 
These analytical results are compared with the results obtained with dedicated numerical 
simulations.
\end{abstract}

 \vskip2pc

\maketitle

\section{Introduction}
\label{sec:intro}

Precision measurements  and precision calculations of the statistical properties of the large-scale structure of the universe
is becoming a key topic in observational and theoretical cosmology,
 in particular in the context of dark energy searches.
Within the concordant model of cosmology (see e.g. \cite{1995Natur.377..600O} and now strongly supported by the observations, see for instance \cite{2003ApJS..148..175S}),  the dynamical growth of structure is a priori governed since $z \sim 3000$ by collision-less dark matter components (although dark energy dominates the  expansion at about $z<1$),  and structures are thought to emerge out of primordial nearly scale-invariant Gaussian metric perturbations. This gives rise to a hierarchical evolution, as increasingly larger scales turn nonlinear (very large scales being in the linear Gaussian regime and small scales in the highly nonlinear regime). Today, scales beyond $\sim 10$ Mpc are well described by linear theory while scales below $\sim 1$ Mpc  are within the highly nonlinear regime. Providing theoretical insights into such a field is therefore a challenging idea.

The highly nonlinear regime has proved very difficult to handle by analytical tools so far (see for instance \cite{2002PhR...367....1B}),
and one must resort to numerical simulations and phenomenological models 
(such as the halo model \cite{2002PhR...372....1C}) that involve some free parameters 
that are fitted to numerical results.

The quasilinear regime on the other hand provides us with a priori 
robust and controlled predictions. In this regime indeed
perturbation theory techniques can be employed and many have already been developed. They are various extensions of the standard Perturbation Theory \cite{2002PhR...367....1B}, such as the RPT \cite{2006PhRvD..73f3519C,2006PhRvD..73f3520C,2008PhRvD..78j3521B}, large-$N$
expansions \cite{2007A&A...465..725V,2008A&A...484...79V},
the closure theory \cite{2008ApJ...674..617T,2009PhRvD..79j3526H}, the time renormalization equation \cite{2008JCAP...10..036P}, etc., usually developed in Eulerian space but sometimes in Lagrangian space as in \cite{2011arXiv1105.1491O}.  
The validity regime of such perturbation theories is however limited in nature, and in a non-controllable way, by the effects of shell crossings. Indeed, as small scale nonlinear regions form, fluid flows originating from different parts of the universe cross each-other leading to local multi-valued velocities, hence to vorticity and effective anisotropic pressure (see review \cite{2002PhR...367....1B}).  So far those effects have been observed to have limited effects on large scales, either through simple analytic investigation, as in \cite{1999A&A...343..663P,2011A&A...526A..67V}, or a posteriori from the successes of perturbation theory results. 
It remains that being able to explore the analytical properties of tools of interest in regimes were shell crossing is present is of crucial importance. 

The Geometrical Adhesion Model (GAM) provides for such an appealing framework \cite{1989MNRAS.236..385G,1994A&A...289..325V}. It is based on the Burgers equation~\cite{0302.60048} in the inviscid limit which, before shell crossing, reproduces the well-known Zel'dovich approximation \cite{1970A&A.....5...84Z}. At shell crossing however particles are prevented from crossing with the introduction of an infinitesimally small viscosity.
The GAM is based on a subsequent prescription concerning the way matter flows within these critical regions. This model is examined in \cite{2010PhRvE..82a6311B,2011PhRvD..83d3508V}
where it is argued that it provides an attractive toy model for the cosmic matter distribution. In this paper we aim at taking advantage of this fact to explore statistical indicators  in their full complexity.  

The quantities we are more particularly interested in are the so-called propagators. 
Indeed, while standard perturbation theory only involves many-body density and
velocity correlation functions, or polyspectra, most resummation schemes 
also involve ``propagators'' or ``response functions''. These unequal-time
quantities describe the evolution with time of small fluctuations, and
often appear at intermediate stages in these resummation procedures.
Such quantities appear for instance in the ``large-$N$'' expansions developed in
\cite{2007A&A...465..725V,2007A&A...476...31V,2008A&A...484...79V}, the ``closure
approximation'' of \cite{2008ApJ...674..617T,2009PhRvD..79j3526H} and also in the ``RPT'' approach of
\cite{2006PhRvD..73f3519C,2006PhRvD..73f3520C,2008PhRvD..78j3521B} where they were initially introduced.
In addition to encapsulating diagrammatic resummations, they also have
a well-defined physical meaning. Propagators are defined, as is usual in statistical physics, as two-time response
functions of the system. For instance the density propagator in real space 
coordinates, $\pR_{\delta\zeta}(\vx_2,t_2;\vx_1,t_1)$, is the ensemble average, over all intervening modes, 
of the functional derivative
\beq
t_2\geq t_1 : \;\; \pR_{\delta\zeta}(\vx_2,t_2;\vx_1,t_1) = 
\left\langle \frac{\cD \delta(\vx_2,t_2)}{\cD\zeta(\vx_1,t_1)} \right\rangle,
\label{Rt2t1}
\eeq
where $\zeta$ is an infinitesimal external perturbation to the density contrast, or velocity field,
applied at time $t_1$. This gives the ``linear response'' of the system, with
respect to the density contrast at position $\vx_2$ and time $t_2$, to an external
perturbations applied at position $\vx_1$ at the earlier time $t_1$~\footnote{Here 
``linear'' refers to the fact that we only consider the response up to linear order over $\zeta$, i.e. for infinitesimal external 
perturbations. However,
$\pR_{\delta\zeta}$ includes the fully nonlinear dynamics of the physical field $\delta$ itself.}.
In this article, we shall focus on the propagator from initial-time fluctuations,
$t_1\rightarrow 0$. Then, this is the response to the initial conditions themselves,
which can be defined through the initial velocity potential $\psi_0$ or the linear
density contrast $\delta_{L0}$.
Thus, as in \cite{2010PhRvD..81d3516B}, we define the density propagator,
\beq
t \geq 0 : \;\; \pR_{\delta}(\vx,t;\vq) = 
\left\langle \frac{\cD \delta(\vx,t)}{\cD\delta_{L0}(\vq)} \right\rangle .
\label{Rt-Eul}
\eeq
It gives the change in the density contrast at position $\vx$ and time $t$
induced by an infinitesimal change to the initial (or linear) density contrast
at position $\vq$.

One of the aims of this paper is to derive the properties of propagators in as much details as possible 
and compare the results with numerical experiments. In particular we want to see how these functions behave, and what they
are sensitive to, when a full nonlinear evolution of the fields is taken into account. There are a priori two 
means of describing the large-scale structure dynamics and properties; one is based on the use of the 
Eulerian coordinates, this is a natural choice since it is directly related to observations, the other is based
on the Lagrangian coordinates. As we shall see below, and as pointed out in
\cite{2007A&A...476...31V,2008PhRvD..78h3503B}, the Eulerian response functions and
propagators are dominated by a ``sweeping effect'', that is, the collective
transport of small-scale density fluctuations by the long wavelengths of the
velocity field~\footnote{at least in the case of single pressureless fluid. This is not the case for multiple fluids, see~\cite{2011arXiv1109.3400B}.}. This holds for both the gravitational and the Burgers dynamics,
and it means that the properties of these two-time functions are not a very good
probe of the properties of the large-scale structures, observed at a given time.
This shortcoming led to the study in \cite{2008PhRvD..78h3503B,2010PhRvD..81d3516B}
of Lagrangian-space propagators. Indeed, in a Lagrangian framework, where one
follows the motion of particles, the impact of uniform translations is automatically
removed (because they do not change the system as viewed from a particle).
Then, propagators or correlation functions automatically go beyond the 
``sweeping effect'' and directly probe the deformation of the density field
(e.g., tidal effects).
This suggests that Lagrangian propagators should generally provide more sensitive
probes of the matter distribution. This is one of the motivations for the study presented here.
In the context of the GAM
it is indeed possible to relate propagators in Lagrangian coordinates to geometrical properties of the late-time
field.

The plan of the paper is the following. In Sect.~\ref{Burgers-dynamics} we present the equations governing
the Burgers dynamics and the geometrical adhesion model. Analytical results regarding this model are 
presented in  the following section \ref{Analytical-results}. In Sect.~\ref{Numerical} we present numerical results that illustrate our 
anaytical findings.

\section{Burgers dynamics and Geometrical adhesion model}
\label{Burgers-dynamics}

\subsection{Equations of motion and geometrical constructions}
\label{eq-motion}

We first recall in this section the definition of the ``geometrical adhesion model'', GAM, described in more details in a previous
paper \cite{2010PhRvE..82a6311B}. This model coincides with the well-known Zel'dovich dynamics \cite{1970A&A.....5...84Z} before shell crossing \cite{1989MNRAS.236..385G,1994A&A...289..325V}, and beyond shell crossing it is built 
from the $d$-dimensional Burgers equation \cite{0302.60048} for the velocity field $\vu(\vx,t)$,
\beq
\pl_t \vu + (\vu\cdot\nabla)\vu = \nu \Delta \vu , \hspace{1cm} \nu\rightarrow 0^+ ,
\label{Burgers}
\eeq
where we consider the inviscid limit.
As is well known, for curlfree initial velocity fields the nonlinear Burgers equation
(\ref{Burgers}) can be solved
through the Hopf-Cole transformation \cite{0039.10403,0043.09902}, by making the
change of variable $\psi(\vx,t)=2\nu\ln\Xi(\vx,t)$, where $\psi(\vx,t)$ is the
velocity potential defined by
\beq
\vu(\vx,t) = - \nabla \psi .
\label{u-psi}
\eeq
This yields the linear heat
equation for $\Xi(\vx,t)$, which leads to the solution
\beq
\psi(\vx,t) = 2\nu\ln\int\frac{\dd\vq}{(4\pi\nu t)^{d/2}} \, 
\exp\left[ \frac{\psi_0(\vq)}{2\nu}-\frac{|\vx-\vq|^2}{4\nu t}\right] .
\label{Hopf1}
\eeq
Then, in the inviscid limit $\nu\rightarrow 0^+$, a steepest-descent method
gives \cite{0302.60048,2007PhR...447....1B}
\beq
\psi(\vx,t) = \sup_{\vq}\left[\psi_0(\vq)-\frac{|\vx-\vq|^2}{2t}\right] .
\label{psixpsi0q}
\eeq
If there is no shock, the maximum in (\ref{psixpsi0q}) is reached at a unique
point $\vq(\vx,t)$, which is the Lagrangian coordinate of the particle that is
located at the Eulerian position $\vx$ at time $t$ 
(hereafter, we note by the letter $\vq$ the Lagrangian coordinates, i.e. the initial
positions at $t=0$ of particles, and by the letter $\vx$ the Eulerian coordinates
at any time $t>0$).
If there are several degenerate solutions to (\ref{psixpsi0q}), we have a shock
at position $\vx$ and the velocity is discontinuous.

Thus, a key property of the nonlinear equation (\ref {Burgers}) is that we know
its explicit solution (\ref{psixpsi0q}) at any time $t$. Therefore, we can build the velocity field
at any time $t$ from the maximization (\ref{psixpsi0q}), which also corresponds
to a Legendre transform, without solving the dynamics at intermediate times.
In a cosmological context we are more particularly interested in the matter
distribution; it is therefore very useful to extend this property to the density field.
As explained in detail in~\cite{2010PhRvE..82a6311B}, this is possible
provided we use a specific continuity equation, which differs from the
usual continuity equation by a peculiar diffusive term, proportional to $\nu$,
that vanishes outside of shocks in the inviscid limit.
More precisely, one obtains in this case the matter distribution from the
Lagrangian map (i.e. the displacement field), $\vq\mapsto\vx$,
which is defined as follows.

Defining the ``linear'' Lagrangian potential $\varphi_L(\vq,t)$ by
\beq
\varphi_L(\vq,t) = \frac{|\vq|^2}{2} - t \psi_0(\vq) ,
\label{phiLdef}
\eeq
so that in the linear regime the Lagrangian map is given by
\beq
\vx_L(\vq,t) =  \frac{\pl\varphi_L}{\pl\vq} = \vq + t \vu_0(\vq) ,
\label{xL}
\eeq
and introducing the function
\beq
H(\vx,t) = \frac{|\vx|^2}{2} + t \psi(\vx,t) ,
\label{Hdef}
\eeq
one can see that the maximum (\ref{psixpsi0q}) can be written as the Legendre
transform (see note~\footnote{
Here we used the standard definition of the Legendre-Fenchel conjugate $f^*(\vs)$
of a function $f(\vx)$, 
$
f^*(\vs) \equiv \cL_{\vs} [ f(\vx) ] = \sup_{\vx} [ \vs\cdot\vx - f(\vx) ]
$.
})
\beq
H(\vx,t) = \sup_{\vq} \left[ \vx\cdot\vq - \frac{|\vq|^2}{2} + t \psi_0(\vq) \right]
= \cL_{\vx} [ \varphi_L(\vq,t) ] .
\label{Hxphiq}
\eeq
Therefore, Eq.(\ref{Hxphiq}) yields the inverse Lagrangian map, $\vx \mapsto \vq$,
$\vq(\vx,t)$ being the point where the maximum in Eq.(\ref{psixpsi0q}) or
(\ref{Hxphiq}) is reached. This is only a rewritting of the Hopf-Cole solution
(\ref{psixpsi0q}), but this is not sufficient to fully define the density field
in dimension greater than one. Indeed, because of shocks there is no unique  
way to invert the mapping $\vx \mapsto \vq$.
Then, the ``geometrical adhesion model'' \cite{1994A&A...289..325V,2010PhRvE..82a6311B}
consists in choosing the direct
Lagrangian mapping, $\vq\mapsto\vx$, as the Legendre conjugate of
$\vx \mapsto \vq$,
\beq
\vq(\vx,t) = \frac{\pl H}{\pl\vx} , \;\;\; \vx(\vq,t) =  \frac{\pl\varphi}{\pl\vq} ,
\label{qx-xq}
\eeq
where the potential $\varphi$ is given by
\beq
\varphi(\vq,t) \equiv \cL_{\vq} [ H(\vx,t) ] = \sup_{\vx} \left[ \vq\cdot\vx 
-  H(\vx,t) \right] .
\label{varphicdef}
\eeq
From standard properties of the Legendre transform, this only gives back the
linear Lagrangian potential $\varphi_L(\vq,t)$ of Eqs.(\ref{phiLdef}) and (\ref{Hxphiq}) if the latter is
convex, and in the general case it gives its \textsl{convex hull},
\beq
\varphi = {\rm conv}(\varphi_L) .
\label{phi-convex}
\eeq

Then, the density field is determined by the conservation of matter
\cite{0753.76004,0919.60004,2010PhRvE..82a6311B}, 
\beq
\rho(\vx,t) \dd\vx = \rho_0 \dd\vq ,
\eeq
which reads as
\beq
\frac{\rho(\vx)}{\rho_0} = \det\left(\frac{\pl\vq}{\pl\vx}\right) 
= \det\left(\frac{\pl\vx}{\pl\vq}\right)^{-1} .
\label{rhoJacob}
\eeq
Here we used the fact that both determinants are positive, thanks to the convexity
of $H(\vx)$ and $\varphi(\vq)$, and we assumed a uniform initial density
$\rho_0$ at $t=0$.
Thus, the ``Lagrangian-Eulerian'' mapping $\vq\leftrightarrow\vx$ of Eq.(\ref{qx-xq})
and the density field (\ref{rhoJacob}) define the ``geometrical adhesion model'' that we
study in this article, where both the velocity and density fields can be obtained at
any time from the initial fields by geometrical constructions.

\subsection{Initial conditions and self-similarity}
\label{Initial-conditions}

As in \cite{2011PhRvD..83d3508V}, we consider Gaussian random initial conditions,
which are isotropic and homogeneous, with power-law initial power spectra.
They can be defined in terms of the initial velocity potential,
$\psi_0(\vx)$, or equivalently in terms of the linear density contrast,
$\delta_L(\vx,t)$, which is the usual approach in cosmology.
Thus, introducing the density contrast,
\beq
\delta(\vx,t) = \frac{\rho(\vx,t)-\rho_0}{\rho_0} ,
\label{deltadef}
\eeq
one can see that in the linear regime \footnote{This amounts to linearize the equation
of motion (\ref{Burgers}) and the continuity equation (which holds before the
formation of shocks).}, where the particles follow the trajectories
(\ref{xL}) that still coincide with the Zel'dovich dynamics \cite{1970A&A.....5...84Z},
the density contrast behaves at linear order as
\beq
\delta_L(\vx,t) = t \, \delta_{L0}(\vx)  , \;\; \mbox{with} \;\; 
\delta_{L0} = -\nabla \cdot \vu_0 =  \Delta \psi_0 .
\label{linear}
\eeq
Going to Fourier space, with the normalization,
\beq
\delta_L(\vx,t) = \int\dd\vk \; e^{\ii\vk \cdot \vx} \; \tdelta_L(\vk,t) ,
\label{Fourier}
\eeq
the linear density field $\delta_L$ is taken as Gaussian, homogeneous, and 
isotropic. Then, it is fully described by its power spectrum
$P_{\delta_L}(k,t)$,
\beq
\lag\tdelta_L\rag=0 , \;\;\; \lag\tdelta_L(\vk_1)\tdelta_L(\vk_2)\rag = 
\delta_D(\vk_1+\vk_2) P_{\delta_L}(k_1) ,
\label{Pdelta-def}
\eeq
which we choose of the power-law form
\beq
-3<n<1 : \;\;\; P_{\delta_L}(k,t) = \frac{D}{(2\pi)^d} \,t^2 \, k^{n+3-d} ,
\label{PdeltaL}
\eeq
where $D$ and $n$ are the amplitude and slope parameters.

In the inviscid limit, one can check \cite{1997JFM...344..339G,2009PhRvE..80a6305V,2011PhRvD..83d3508V}
that the power-law initial conditions (\ref{PdeltaL})
give rise to a self-similar dynamics~\footnote{This only holds in the range $-3<n<1$.}:
a rescaling of time is statistically equivalent to a rescaling of distances, as
\beq
\lambda>0: \;\; t\rightarrow \lambda t, \;\; \vx \rightarrow 
\lambda^{2/(n+3)} \vx .
\label{selfsimilar}
\eeq
Thus, the system displays a hierarchical evolution and the only characteristic
scale at a given time $t$ is the so-called integral scale of turbulence, $L(t)$,
which is generated by the Burgers dynamics and grows with time as in 
(\ref{selfsimilar}). Hereafter we choose the normalization
\beq
L(t) =  (2Dt^2)^{1/(n+3)} ,
\label{Lt}
\eeq
where the constant $D$ was defined in Eq.(\ref{PdeltaL}).
This scale measures the typical distance between shocks,
and it separates the large-scale quasi-linear regime, where the density power
spectrum keeps its initial power-law form, (\ref{PdeltaL}), from the small-scale
nonlinear regime, which is governed by shocks
and pointlike masses, where the density power spectrum reaches the universal
white-noise behavior (i.e. $P_{\delta}(k,t)$ has a finite limit for $k\gg 1/L(t)$).
Detailed numerical studies of the cluster mass functions and density fields generated
in 1D and 2D are presented in \cite{1994A&A...289..325V,2011PhRvD..83d3508V}.

\section{Analytical results}
\label{Analytical-results}

\subsection{Definitions}

\subsubsection{Propagators and correlators, general properties}

As noticed in \cite{2006PhRvD..73f3520C} from the perturbative expansion
of the nonlinear density contrast $\delta(\vx,t)$ over powers of the linear growing
mode $\delta_{L0}$, for Gaussian initial conditions the response function (\ref{Rt-Eul}) 
is related to the cross-correlation $\xi_{\delta}(\vx,t;\vq)$ by
\beqa
\xi_{\delta}(\vx,t;\vq)&\equiv& \langle \delta(\vx,t)\delta_{L0}(\vq)\rangle\nonumber\\
&=&\int \dd \vq' \, \pR_{\delta}(\vx,t;\vq') \langle \delta_{L0}(\vq')\delta_{L0}(\vq)\rangle .
\label{S-def}
\eeqa
Following the approach used in \cite{2007A&A...476...31V}, this relation can actually be obtained by
a simple integration by parts and does not explicitly rely on perturbative expansions.
Indeed, writing the nonlinear density contrast as a functional of the initial
condition $\delta_{L0}$,
\beq
\delta(\vx,t) = \cF_{\vx,t}[\delta_{L0}(\vq)] ,
\label{F-def}
\eeq
the response function (\ref{Rt-Eul}) writes as
\beq
\pR_{\delta}(\vx,t;\vq) = \int\cD \delta_{L0} \;  
e^{-\frac{1}{2} \delta_{L0} \cdot C_0^{-1} \cdot \delta_{L0}} \;
\frac{\cD \cF_{\vx,t}[\delta_{L0}]}{\cD\delta_{L0}(\vq)} \;  ,
\label{R-funct}
\eeq
where we did not write an irrelevant normalization constant and
$C_0(\vq_1,\vq_2)=\lag \delta_{L0}(\vq_1) \delta_{L0}(\vq_2) \rag$
is the two-point correlation of the initial Gaussian field.
Integrating Eq.(\ref{R-funct}) by parts gives
\beqa
\pR_{\delta}(\vx,t;\vq) & = & \!\! \int\cD \delta_{L0} \;  
e^{-\frac{1}{2} \delta_{L0} \cdot C_0^{-1} \cdot \delta_{L0}} \;
\cF_{\vx,t}[\delta_{L0}] \nonumber \\
&& \times  \left[ \int \dd \vq' \, C_0^{-1}(\vq,\vq')\cdot\delta_{L0}(\vq') \right]
\\
& = & \!\! \int \dd \vq' \, C_0^{-1}(\vq,\vq')\cdot \lag \delta(\vx,t) \delta_{L0}(\vq') \rag .
\label{R-funct-2}
\eeqa
Multiplying by the operator $C_0$ we recover Eq.(\ref{S-def}). This result obtained here
for Gaussian initial conditions can obviously be extended to non-Gaussian initial conditions
(but then the integration by parts gives rise to a more complicated functional form of
the cross-correlation with the initial fields).
What this simple calculation shows is that the identities (\ref{S-def}) and (\ref{R-funct-2}) are quite general. They
apply to non-analytic functionals $\cF_{\vx,t}$ and should be valid as soon as propagators actually exist. 
In particular, these relations remain valid after
shell crossing, where perturbative expansions break down. 
This is confirmed in App.~\ref{Cross-correlations} where we do not detect
any deviation from the identities  (\ref{S-def}) and (\ref{R-funct-2}) far in the
nonperturbative regime.
The full validity of this identity was overlooked in
early papers but was  successfully checked in numerical simulations in  \cite{2006PhRvD..73f3520C}.
This is a useful property since it is usually more
convenient to measure cross-correlations such as $\lag \delta(\vx,t) \delta_{L0}(\vq)\rag$ in numerical simulations
whereas, in the case we consider here, the expression (\ref{Rt-Eul}) is better suited to analytical investigations.

\subsubsection{Eulerian quantities}

In the following we will be interested in quantities defined in both Eulerian coordinates and Lagrangian 
coordinates. In practice, since the system is statistically homogeneous and isotropic, it is
convenient to work in Fourier space, and we define the Fourier-space counterpart
of the response $G_{\delta}$ of Eq.(\ref{Rt-Eul}) by
\beq
t\geq 0 : \;\; \left\langle\frac{\cD\tdelta(\vk,t)}{\cD\tdelta_{L0}(\vk')}\right\rangle =
\delta_D(\vk-\vk') \, \tG_{\delta}(k,t) ,
\label{tR-def}
\eeq
which is related to the real-space response by
\beq
\tG_{\delta}(k,t)  = \int \dd\vx \, e^{-\ii\vk\cdot\vx} \, \pR_{\delta}(x,t) ,
\label{tR-R}
\eeq
where we used $\pR_{\delta}(\vx,t;\vq)=\pR_{\delta}(|\vx-\vq|,t)$.
The relation (\ref{S-def}) reads in Fourier space as
\beq
\left\langle\tdelta(\vk,t)\tdelta_{L0}(\vk')\right\rangle = \delta_D(\vk+\vk') \, 
\tG_{\delta}(k,t) \, P_{L0}(k) .
\label{tP-def}
\eeq
Here we focused on the density propagator or response function, but we can
also consider the velocity potential $\psi$,
such as
\beq
\pR_{\psi}(\vx,t;\vq) = \left\langle\frac{\cD\psi(\vx,t)}{\cD\psi_0(\vq)}\right\rangle ,
\label{Rpsi-def}
\eeq
and mixed statistics, such as
\beq
\pR_{\delta\psi}= \left\langle\frac{\cD\delta(\vx,t)}{\cD\psi_0(\vq)}\right\rangle ,
\label{Rdeltapsi-def}
\eeq
and their respective Fourier-space counterparts.

\subsubsection{Lagrangian quantities}
\label{Lagrangian-prop}

Similar quantities can be defined in a Lagrangian framework.
Following \cite{2008PhRvD..78h3503B} we first introduce the divergence $\kappa$ of the displacement field,
\beq
\kappa(\vq,t) = d - \frac{\pl\vx}{\pl\vq} = - \nabla_{\vq} \cdot [ \vx(\vq,t)-\vq ] ,
\label{kappa-def}
\eeq
where $\vx(\vq,t)$ is the position at time $t$ of the particle of Lagrangian
coordinate $\vq$. From the conservation of matter, Eq.(\ref{rhoJacob}), we can
see that at linear order $\kappa$ is also the density contrast,
\beq
\kappa_L(\vq,t) = \delta_L(\vq,t) = t \, \kappa_{L0}(\vq)  \hspace{0.3cm}
\mbox{with} \hspace{0.3cm} \kappa_{L0}(\vq) = \delta_{L0}(\vq) .
\label{kappaL}
\eeq
It is clear from the definition (\ref{kappa-def}) that a uniform translation
does not change the value of $\kappa$, so that it is not sensitive to the
``sweeping effect'' discussed in the introduction 
\cite{2007A&A...476...31V,2008PhRvD..78h3503B}. Then, as above,
we define the associated Lagrangian propagators,
\beq
t \geq 0 : \;\; \pR_{\kappa\psi}(\vq,t;\vq_0) = 
\left\langle \frac{\cD \kappa(\vq,t)}{\cD\psi_{L0}(\vq_0)} \right\rangle ,
\label{Rt-Lag1}
\eeq
and
\beq
t \geq 0 : \;\; \pR_{\kappa}(\vq,t;\vq_0) = 
\left\langle \frac{\cD \kappa(\vq,t)}{\cD\kappa_{L0}(\vq_0)} \right\rangle .
\label{Rt-Lag2}
\eeq
This again reads in Fourier space as
\beq
t\geq 0 : \;\; \left\langle\frac{\cD\tkappa(\vk,t)}{\cD\tpsi_{L0}(\vk')}\right\rangle =
\delta_D(\vk-\vk') \, \tG_{\kappa\psi}(k,t) ,
\label{tR-def-Lag1}
\eeq
and
\beq
t\geq 0 : \;\; \left\langle\frac{\cD\tkappa(\vk,t)}{\cD\tkappa_{L0}(\vk')}\right\rangle =
\delta_D(\vk-\vk') \, \tG_{\kappa}(k,t) ,
\label{tR-def-Lag2}
\eeq
Using the property $\tkappa_{L0}(k) = \tdelta_{L0}(k) = - k^2 \tpsi_0(k)$,
from Eqs.(\ref{kappaL},\ref{linear}), we also have the relation
\beq
\tR_{\kappa}(k,t)= - k^{-2} \, \tR_{\kappa\psi}(k,t) .
\label{Rkappa-Rkappapsi}
\eeq
Similarly to the Eulerian case we also have the relation
\beq
\left\langle\tkappa(\vk,t)\tkappa_{L0}(\vk')\right\rangle = \delta_D(\vk+\vk') \, 
\tG_{\kappa}(k,t) \, P_{L0}(k) .
\label{tS-def-Lag}
\eeq
Thanks to the explicit Hopf-Cole solution (\ref{psixpsi0q}) and its geometrical
interpretation in terms of first-contact parabolas
\cite{0302.60048,2007PhR...447....1B},
very fruitful insights on the behavior of the propagators can be obtained.
Exact results for the one-dimensional case have been recently derived in
\cite{2010PhRvD..81d3516B}, which we will briefly recall in the course of the general calculation.

\subsection{Eulerian response functions}
\label{Eulerian-propagators}

To take advantage of the Hopf-Cole solution (\ref{psixpsi0q}) it is convenient to
first consider the response function $\pR_{\psi}$ as defined in Eq.(\ref{Rpsi-def}). 

In any dimension we can use the Hopf-Cole solution (\ref{Hopf1})
to obtain the propagator $\pR_{\psi}$. Thus, as in the 1D case
\cite{2010PhRvD..81d3516B}, taking the functional derivative of Eq.(\ref{Hopf1})
and next taking the inviscid limit we obtain
\beqa
\pR_{\psi}(\vx,t;\vq_0) & = & \langle \delta_D[\vq(\vx,t)-\vq_0] \rangle 
= p_{\vx}(\vq_0,t) \nonumber \\
& = &\frac{1}{t^d} \, p(\vu,t) ,
\label{Rpsi-pxu}
\eeqa
with
\begin{equation}
\vu=\frac{\vx-\vq_0}{t}.
\end{equation}
Here $p_{\vx}(\vq_0,t)$ is the probability distribution function of the Lagrangian coordinate
$\vq_0$ of the particle that is located at position $\vx$ at time $t$, and $p(\vu,t)$
is the probability distribution function of the velocity $\vu$ at position $\vx$ and time
$t$ (it does not depend on $\vx$ because of statistical homogeneity).

Therefore, the Eulerian propagator $\pR_{\psi}$ is always given by the one-point
velocity probability distribution, irrespectively of the space dimension. 
As such, it is still governed by the ``sweeping effect'' mentioned in the introduction, and for convergent initial power spectra it obeys
in the weakly nonlinear regime behavior, 
\beqa
\tR^{{\rm RPT}}_{\psi}(k,t) &\sim& t \,  e^{-t^2k^2\sigma^2_{u_0}/2} ,\nonumber \\
\pR^{{\rm RPT}}_{\psi}(\vx,t;\vq) &\sim &\frac{1}{(\sqrt{2\pi}t\sigma_{u_0})^d} \, 
e^{-|\vx-\vq|^2/(2t^2\sigma^2_{u_0})} , 
\label{R-wnl}
\eeqa
where $\sigma^2_{u_0}=\langle |\vu_0|^2\rangle/d$ is the variance of the initial
velocity along any given direction. The behavior (\ref{R-wnl}), which is more general
than the adhesion model, also applies to the gravitational dynamics in general. Indeed, it only
relies on the transport of particles by long-wavelength modes of the velocity field (see 
\cite{2007A&A...476...31V,2011arXiv1109.3400B} for an account of this effect in a perturbative approach),
in other words, on the Galilean invariance further discussed in Sect.~\ref{sum-rules}
below. It is also at the basis of the RPT resummation scheme introduced in
cosmology \cite{2006PhRvD..73f3519C}.

As discussed in \cite{2010PhRvD..81d3516B} for $d=1$, and remains valid in higher
dimensions, for the power-law initial conditions (\ref{PdeltaL})
the weakly nonlinear regime (\ref{R-wnl}) does not exist. 

For $-3<n<-1$, the variance
$\sigma^2_{u_0}$ diverges because of low-$k$ modes. This means that as we
increase the size of the system (since in practice, for instance in numerical
simulations, we consider finite-size systems and eventually take the infinite-size
limit) the contribution from long wavelengths keeps increasing and particles
are transported over increasingly large distances. This yields a divergent
``sweeping effect'' and the propagators vanish as soon as $t>0$.
However, this is not a genuine loss of memory since this divergence
is due to almost uniform random translations and thanks to Galilean invariance
the structures of the density field are not affected. 
In particular, as we recall in Sect.~\ref{Lagrangian-propagators} below, Lagrangian
propagators remain finite.

For  $-1<n<1$, the variance $\sigma^2_{u_0}$ diverges because of
high-$k$ modes. However, as soon as $t>0$ this ultraviolet divergence is
regularized by the dynamics, more precisely by the nonperturbative
``sticking'' of the particles at collision. This means that the one-point
velocity probability distribution $p(\vu,t)$, whence the response functions,
are finite and well-defined. However, they are governed by nonperturbative
effects and do not obey the Gaussian behavior (\ref{R-wnl}).
For $d=1$ and in the case $n=0$ one can actually obtain its exact expression
\cite{2000JFM...417..323F,2009JSP...137..729V,2010PhRvD..81d3516B}, and this gives
in particular the exponential decays $\pR \sim e^{-(x-q)^3/t^2}$ and
$\tR \sim e^{-t k^{3/2}}$ at large $|x-q|$ and large $k$.
For generic $n$ we only know the large-separation tail
\cite{0944.60073,0897.60075,2009PhRvE..80a6305V}
\beq
-1<n<1 , \;\; |x-q|  \rightarrow \infty : \;\;\; \pR_{\psi}(x,t;q) \sim e^{-|x-q|^{n+3}/t^2} ,
\label{REul_asymp}
\eeq
which depends on $n$, contrary to the weakly nonlinear regime behavior
(\ref{R-wnl}).
For $d>1$ we no longer have exact expressions for $p(\vu,t)$, even for $n=0$, but 
in the large-separation limit we expect from rare-event analysis
\cite{2009PhRvE..80a6305V} the tail
\beq
-1<n<1 , \;\; |\vx-\vq| \rightarrow \infty : \;\; \pR_{\psi}(\vx,t;\vq) 
\sim e^{-|\vx-\vq|^{(n+3)}/t^2} ,
\label{Rpsi-d-tail}
\eeq
as in (\ref{REul_asymp}).

The probability distributions $p(\vq_0,t)$ and $p(\vu,t)$ being well defined and normalized
to unity for $-1<n<1$, we obtain from Eq.(\ref{Rpsi-pxu}) the low-$k$ limit
\beq
\tR_{\psi}(0,t) = 1 .
\label{R-psi-k0}
\eeq
This agrees with the linear regime prediction but this exact result is more general since it follows
from the nonperturbative identity (\ref{Rpsi-pxu}). Thus, it remains
valid for the scale-invariant initial conditions with $-1<n<1$, where there is no true weakly nonlinear
regime in the sense of (\ref{R-wnl}) because linear velocity fields are divergent.
However, particles are not transported
arbitrarily far away but on distances that scale with the characteristic length
$L(t)$ defined in (\ref{Lt}).
Thus, even though the ``confining'' process is
strongly nonperturbative, due to the collisions and sticking of particles,
initial density fluctuations seen on scales much larger than $L(t)$ are not erased
and evolve as in linear theory (in a weak sense, that is, if we consider smooth
windows such as a Gaussian rather than a top-hat).
In terms of Burgers turbulence this is related to the principle of ``permanence 
of large eddies'' \cite{1997JFM...344..339G}.

For $d=1$ the density contrast $\delta(x,t)$ associated
with the matter distribution (\ref{rhoJacob}) also reads as
\beq
\delta(x,t) = t \, \frac{\pl^2\psi}{\pl x^2} .
\eeq
Then, from Eq.(\ref{Rpsi-pxu}) one obtains (e.g., by going to Fourier space)
\beq
\pR_{\delta}(x,t;q) =  p(u,t)  , \;\;\; \mbox{with} \;\;\;
u=\frac{x-q}{t} .
\label{Rdelta-pxu}
\eeq
In agreement with physical arguments and perturbative analysis
\cite{2007A&A...476...31V,2008PhRvD..78h3503B}, the results (\ref{Rpsi-pxu}) and
(\ref{Rdelta-pxu}) explicitly show that Eulerian response functions
are dominated by the ``sweeping
effect'', that is, by the transport of particles by the underlying velocity field.
Since only the one-point velocity distribution appears in (\ref{Rdelta-pxu}),
these Eulerian propagators are not very sensitive probes of the density field
(which depends on relative motions).

For $d\geq 2$ the density contrast is no longer related to the velocity potential
$\psi$ by a linear relationship. From Eqs.(\ref{qx-xq},\ref{Hdef}) we obtain
\beq
\frac{\pl q_i}{\pl x_j} = \delta_{i,j} + t \, \frac{\pl^2\psi}{\pl x_i\pl x_j} ,
\label{qiqj}
\eeq
and Eq.(\ref{rhoJacob}) yields
\beq
1+\delta(\vx,t) = \det\left( \delta_{i,j} + t \, \frac{\pl^2\psi}{\pl x_i\pl x_j} \right) ,
\label{delta-psi-d}
\eeq
which contains terms up to power $d$ over $\psi$.
The Eulerian density response function must still be sensitive to the
``sweeping effect'' and dominated by the properties of the velocity field,
in agreement with the response function of the velocity potential
$\psi$ itself. However, because of the nonlinear dependence (\ref{delta-psi-d})
the response function $\pR_{\delta}$ is no longer merely proportional to the one-point
velocity probability distribution and the relationship is more complex, and depends
on the dimension $d$.
For power-law initial conditions it should again vanish as soon as $t>0$ for
$-3<n<-1$, and be finite for $-1<n<1$, with the asymptotic behavior
(\ref{Rpsi-d-tail}) at very large separations.

\subsection{Lagrangian response functions}
\label{Lagrangian-propagators}

\subsubsection{General formalism}

\begin{figure}
\begin{center}
\epsfxsize=7. cm  \epsfysize=6 cm {\epsfbox{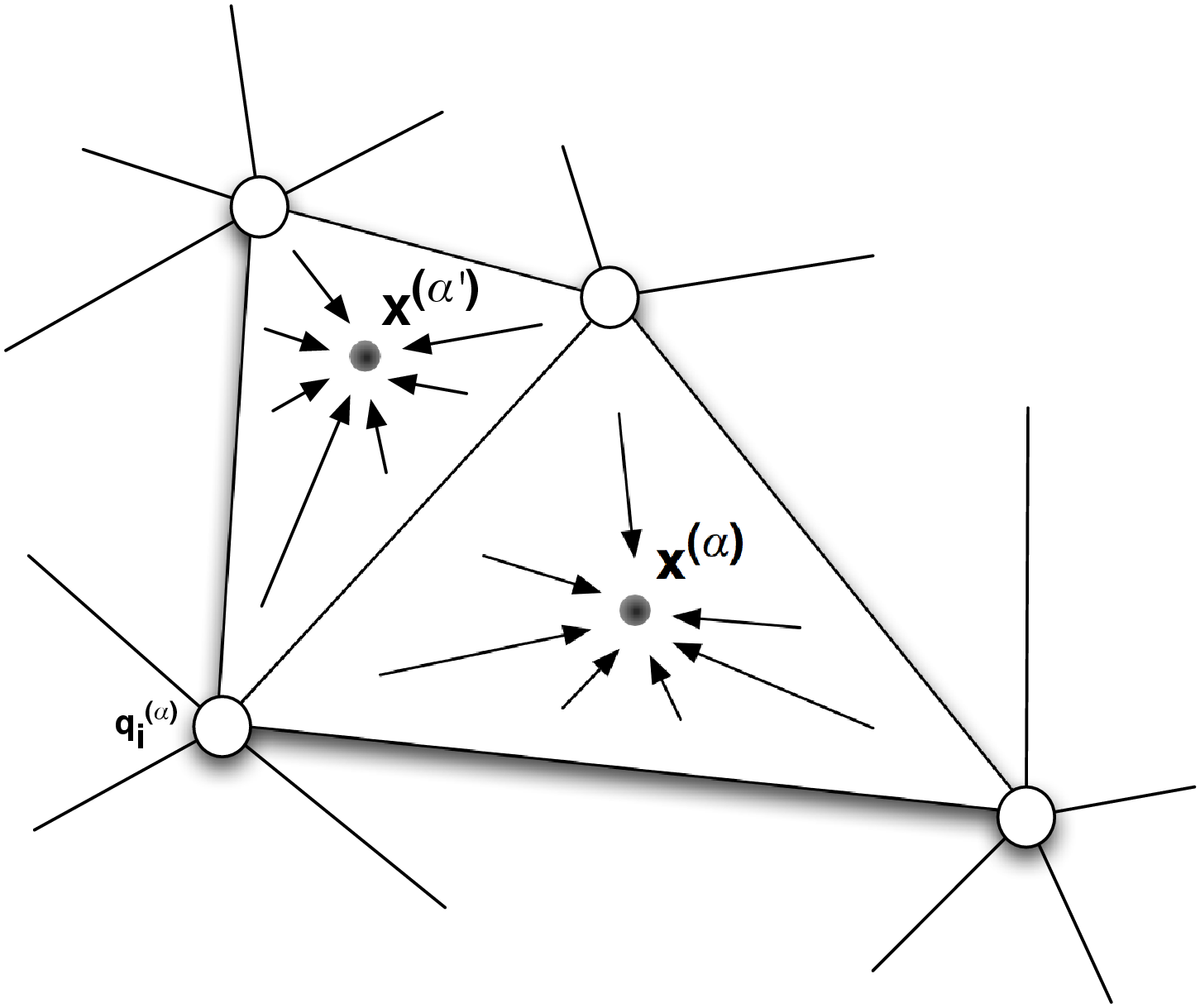}}
\end{center}
\caption{In a triangulation, the matter in each triangle moves toward a single halo. Infinitesimal 
variation of the potential at position $\vq_{i^{(\alpha)}}$
induces an infinitesimal change in the displacement field within the triangles $\vq_{i^{(\alpha)}}$ is a summit of. This change can entirely be described
by a change in the positions of the corresponding halos.}
\label{Triangulation2D}
\end{figure}

As seen in \cite{0755.60104,2010PhRvE..82a6311B},
for the power-law initial conditions that we consider here, all the matter clusters
into a set of point masses that builds a Voronoi-like tessellation in Eulerian
space and a dual triangulation in Lagrangian space.
This Lagrangian-space tessellation is built from segments in 1D, triangles in 2D,
tetrahedra in 3D, and higher-order simplices in higher dimensions \cite{0753.76004}.
A sketch of such a partition in 2D is to be found on Fig. \ref{Triangulation2D}. In this picture one halo is associated with each Lagrangian triangle and it is formed from all the particles that were initially in this corresponding triangle. As a result the mass distribution is nothing but the volume distribution 
of those triangular cells. Moreover, the position of the halos in Eulerian space depends 
on the potential heights $\varphi(\vq_i^{(\alpha)})$ on each of the summits 
$\vq_i^{(\alpha)}$ of the triangle. As time evolves the relative heights of the summits change leading to halo motions and eventually to halo mergings. 
These processes were described in details in  \citep{2010PhRvE..82a6311B}.

The final matter distribution can be characterized by the density field or equivalently by
the displacement field, a quantity that better suits
the Lagrangian description.  What this construction tells us is that within each Lagrangian cell $\cV^{(\alpha)}$ the Lagrangian mapping 
$\vx(\vq)=\vx^{(\alpha)}$ is constant, as all particles that originate from this region belong to a single point-like cluster at position 
$\vx^{(\alpha)}$ at time $t$.
And because the displacement is potential it is entirely determined by its divergence $\kappa$. It is clear here that 
the divergence within each cell is constant. Indeed from the definition (\ref{kappa-def}) this yields $\kappa(\vq,t)=d$ within each
of these Lagrangian cells, with the addition of a Dirac term on the boundaries of
these cells associated with the jump from $\vx^{(\alpha)}$ to $\vx^{(\alpha')}$ as one goes from
cell $\cV^{(\alpha)}$ to cell $\cV^{(\alpha')}$.
This more formally reads as
\beq
\kappa(\vq) = d - \int_{\Sigma} \dd^{d-1} s \; \delta_D[\vq-\vq_{\Sigma}(\vs)]
\; |\vn \cdot \Delta\vx|
\label{kappa-d}
\eeq
where $\vq_{\Sigma}(\vs)$ is the $(d-1)$-dimensional manifold $\Sigma$ built by
the boundaries of the Lagrangian cells (i.e., all triangle sides in 2D),
parameterized by a coordinate $\vs$, $\dd^{d-1} s$ is the natural surface element
on $\Sigma$ embedded in $d$ dimensions, $\vn$ is the
unit normal vector to this manifold and $\Delta \vx = \vx^{(\alpha')}-\vx^{(\alpha)}$ is the separation
vector between the Eulerian positions $\vx^{(\alpha)}$ and $\vx^{(\alpha')}$ of the two mass clusters
associated with the two Lagrangian cells $(\alpha)$ and $(\alpha')$ that are on either side of
$\Sigma$.
In 1D the integral (\ref{kappa-d}) simplifies to the sum over all endpoints $q_i$
of the Lagrangian intervals $[q_i,q_{i+1}]$, associated with the shocks of
Eulerian position $x_i$, as $\kappa(q) = 1 - \sum_i \delta_D(q-q_i) \; (x_i-x_{i-1})$
\citep{2010PhRvD..81d3516B}.

It is convenient to decompose the integral (\ref{kappa-d}) over all cells $\cV^{(\alpha)}$.
The absolute value $|\vn\cdot\Delta\vx|$ arose from the
convexity of the mapping $\vx(\vq)$, that is, the fact that $\varphi(\vq)$ is
convex in Eq.(\ref{qx-xq}). This ensures that $\pl x_i/\pl q_i \geq 0$ along any
direction $i$ 
\cite{0823.76058,2003MNRAS.346..501B,2011A&A...526A..67V}, so that the contribution
$(\vn\cdot\Delta\vx)$ must be taken positive. This can also be written as
\beq
- |\vn \cdot \Delta\vx| = \vn^{(\alpha)}_{\rm out} \cdot \vx^{(\alpha)} +
\vn^{(\alpha')}_{\rm out} \cdot \vx^{(\alpha')} ,
\label{Deltax-a-ap}
\eeq
where $\vn^{(\alpha)}_{\rm out}$ and $\vn^{(\alpha')}_{\rm out}$ are the unit normal
vectors to the surface $\Sigma^{(\alpha,\alpha')}$ that point outward from the neighboring
cells $\cV^{(\alpha)}$ and $\cV^{(\alpha')}$.
Then, Eq.(\ref{kappa-d}) reads as
\beq
\kappa(\vq) = d + \sum_{(\alpha)} \int_{\Sigma^{(\alpha)}} \dd \vs \cdot \vx^{(\alpha)}
\; \delta_D[\vq-\vq_{\Sigma^{(\alpha)}}(\vs)] ,
\label{kappa-d-alpha}
\eeq
where we note $\dd\vs= \vn_{\rm out} \dd s$.

Next, to estimate the Lagrangian response functions (\ref{Rt-Lag1},\ref{Rt-Lag2}) we must consider
the variation of Eq.(\ref{kappa-d-alpha}) for infinitesimal changes of the initial conditions.
The manifold $\Sigma$ is set by the convex hull construction (\ref{phi-convex}).
For instance, in 1D the boundaries $q_i$ are the contact points between the
linear Lagrangian potential $\varphi_L(q)$ and its convex envelope $\varphi$,
whereas in 2D the triangles of the Lagrangian-space tessellation are the triangular
facets of the convex hull $\varphi$ (and the triangle summits $\vq_i^{(\alpha)}$
are again the contact points between $\varphi_L$ and $\varphi$, for each facet
$(\alpha)$ and $i=0,1,2$ indexes the three summits).
Then, in the nondegenerate case (e.g., in 2D a planar facet only makes contact
with three points), which has a unit probability, an infinitesimal change of the
initial conditions, whence of $\varphi_L(q)$, does not modify the set 
$\{\vq_i^{(\alpha)}\}$ of the contact points, but only their heights
$\varphi(\vq_i^{(\alpha)})$. 
This is because for the power-law initial conditions 
(\ref{PdeltaL}) the potentials $\psi_0$ and $\varphi_L$ have no finite second-derivatives
and the contact points are isolated infinitesimally-thin spikes, see 
\citep{2010PhRvE..82a6311B}.
As a result, infinitesimal variations of the initial conditions do not change the manifold
$\Sigma$ neither
 the Lagrangian cells $\cV^{(\alpha)}$; they 
 only change the Eulerian positions $\vx^{(\alpha)}$ of the mass clusters associated with each Lagrangian cell.
 This is the core property which determines how Lagrangian response functions and halo mass functions are related.
We then have
\beq
\frac{\cD\kappa(\vq)}{\cD\psi_0(\vq')} = \sum_{(\alpha)} \int_{\Sigma^{(\alpha)}}
\dd\vs \cdot \frac{\cD \vx^{(\alpha)}}{\cD\psi_0(\vq')}
 \; \delta_D[\vq-\vq_{\Sigma^{(\alpha)}}(\vs)]  .
\label{dkappa-dpsi0-d}
\eeq

Following the definition (\ref{Rt-Lag1}) the Lagrangian propagator is then given by the statistical
average of Eq.(\ref{dkappa-dpsi0-d}).
Because the system is statistically homogeneous and isotropic,
$\pR_{\kappa\psi}(\vq,\vq')$ only depends on $|\Delta\vq|$ with
$\Delta\vq=\vq-\vq'$, and we obtain
\beq
\pR_{\kappa\psi}(\Delta\vq,t) = \int_0^{\infty}\dd m \; n(m) \left\lag \int_{\Sigma} \dd\vs \cdot  \frac{\cD \vx}{\cD\psi_0(\vq(\vs)-\Delta\vq)} \right\rag_{\!m}  
\label{Gkappapsi-nm}
\eeq
where $n(m) \dd m$ is the mean number of Lagrangian cells of mass $m$ (or of point-like
clusters of mass $m$ in Eulerian space), per unit volume.
Here $\lag .. \rag_m$ is the statistical average over cells of fixed mass $m$.
In 1D this is immediate since Lagrangian intervals are fully defined by their mass
(i.e. their length), but in higher dimensions we must
average over the distribution of shapes and angles at fixed mass $m$.
We will show explicit examples of such averaging in the following.

As recalled above, the Eulerian position $\vx$ of the mass cluster only depends on the
values $\psi_{0;i}$ of the velocity potential at the $(d+1)$ summits $\vq_i$ that define 
the Lagrangian cell $\cV$, hence we can write the functional derivative in
Eq.(\ref{Gkappapsi-nm}) as
\beq
\frac{\cD\vx}{\cD\psi_0(\vq')} = \sum_{i=0}^d \frac{\pl\vx}{\pl \psi_{0;i}} \;
\delta_D(\vq'-\vq_i) .
\label{Dx-Dpsi0-sum}
\eeq 
By symmetry the contributions of all $(d+1)$ summits are identical, after we average
over shapes and angles, and choosing the origin of coordinates on summit $\vq_0$
we obtain
\beqa
\pR_{\kappa\psi}(\Delta\vq,t) & = & (d+1) \int_0^{\infty}\dd m \; n(m) \nonumber \\
&& \hspace{-1cm}  \times \left\lag \int_{\Sigma} \dd\vs \cdot  \frac{\pl \vx}{\pl\psi_{0;0}} \; \delta_D(\vq(\vs)-\Delta\vq) \right\rag_{\!m}   ,
\label{Gkappapsi-nm-0}
\eeqa
where $\vq_0=0$.
This yields in Fourier space
\beq
\tR_{\kappa\psi}(k,t) = - t \, k^2  \int_0^{\infty}\dd m \, n(m) \, \ell^d \, \tW(k;m) ,
\label{tRkappapsi-W}
\eeq
where $\ell$ is the typical size of cells of mass $m$, defined as
\beq
m= \rho_0 \ell^d ,
\label{ell-def}
\eeq
and the dimensionless kernel $\tW(k;m)$ reads as
\beq
\tW(k;m) = \frac{-(d+1)}{t k^2 \ell^d} \; \left\lag \int_{\Sigma} \dd\vs \cdot 
\frac{\pl \vx}{\pl\psi_{0;0}} \; e^{-\ii\vk\cdot[\vq(\vs)-\vq_0]} \right\rag_{\!m} ,
\label{W-def}
\eeq
where again $\lag .. \rag_m$ is the statistical average over cells of fixed mass $m$.
Then, from Eq.(\ref{Rkappa-Rkappapsi}) the propagator $\tR_{\kappa}$ writes as
\beq
\tR_{\kappa}(k,t) = t  \int_0^{\infty}\dd m \, n(m) \, \ell^d \, \tW(k;m) .
\label{tRkappa-W}
\eeq

\subsubsection{Large scale behavior}
\label{sum-rules}

We first explore the behavior of $\tW(k;m)$ for $k\to 0$ and show that it correctly reproduces 
what one expects from the linear theory. For that we simply use  Eq.(\ref{W-def})  and Taylor expand it
at $k\approx 0$. Let us first note that the term in $1/k$ vanishes by symmetry, because the factors $\vk\cdot[\vq(\vs)-\vq_i]$
average to zero as we integrate over the global orientation of the Lagrangian cell $\cV$.
We then get

\beqa
\tW(k;m)&=& \frac{-1}{k^2\,t\ell^d} \left\lag \int_{\Sigma} \dd \vs \, \cdot 
\sum_i   \frac{\pl \vx}{\pl\psi_{0;i}} \right\rag_{\!m} \nonumber\\
 &&\hspace{0cm}+\frac{1}{2 t\ell^d} \left\lag \sum_i \int_{\Sigma} \dd\vs \cdot
\frac{\pl \vx}{\pl\psi_{0;i}} \; (\hat{\vk}\cdot[\vq-\vq_i])^2  \right\rag_{\!m} \nonumber\\
&&\hspace{0cm}+\cO(k^2),
\eeqa
where we replaced the factor $(d+1)$ by the sum over the derivatives with respect to the
$(d+1)$ summits, going back to Eq.(\ref{Dx-Dpsi0-sum}), 
and where in the second term $\hat{\vk}=\vk/|\vk|$ is a unit vector along an arbitrary direction.
It can then be observed that the sum in the first term vanishes since the position
$\vx$ of the mass cluster is not modified by a uniform shift of the velocity potential. We can also observe that, 
using Gauss' theorem, the integral 
on the surface $\Sigma$ in the second term can be written in terms of
an integral on the volume $\cV$. Using again the property $\sum_i \pl\vx/\pl\psi_{0;i}=0$,
this yields a volume factor $|\cV|=\ell^d$ so that
\beq
\tW(k;m) = \frac{-1}{t} \left\lag \sum_i (\hat{\vk}\cdot\vq_i) (\hat{\vk}\cdot\frac{\pl \vx}{\pl\psi_{0;i}} ) \right\rag_{\!m} 
+\cO(k^2).
\label{tW-sum-2}
\eeq
This sum is in fact constrained by the Galilean invariance of the dynamics.
Indeed, adding a uniform initial velocity $\vv_0$, $\vu_0(\vq) \rightarrow \vu_0(\vq)+\vv_0$,
also corresponds to the changes
\beq
\vx(\vq,t) \rightarrow \vx(\vq,t)+\vv_0 t , \;\; \psi_0(\vq) \rightarrow \psi_0(\vq)-\vv_0\cdot\vq
\label{Galilean}
\eeq
and does not modify the structure of the Lagrangian-space tessellation.
This implies that an infinitesimal uniform velocity perturbation $\vv_0$ leads to the
shift of cluster position
\beq
\vv_0 t = \Delta \vx = \sum_i \frac{\pl\vx}{\pl\psi_{0;i}} \Delta \psi_{0;i} 
=  - \sum_i \frac{\pl\vx}{\pl\psi_{0;i}} (\vv_0\cdot\vq_i) ,
\eeq
and taking the scalar product with $\vv_0$ yields
\beq
|\vv_0|^2 = \frac{-1}{t} \sum_i (\vv_0\cdot\vq_i) (\vv_0\cdot \frac{\pl\vx}{\pl\psi_{0;i}}) .
\eeq
This equation holds for $\vv_0$ of any direction and any length (since both sides
scale linearly with $|\vv_0|^2$), and taking $|\vv_0|=1$ we obtain that the average in
(\ref{tW-sum-2}) writes as $\tW(0;m) = \lag 1 \rag_m$, whence
\beq
\tW(k;m) =1 +\cO(k^2),
\label{tW-norm}
\eeq
independently of the statistical properties of the Lagrangian-space tessellation.

Finally, since all the matter is contained in the mass clusters, the mass function obeys the
normalization
\beq
\int_0^{\infty} \dd m \, n(m) \, \frac{m}{\rho_0} = 1 ,
\eeq
and using Eqs.(\ref{ell-def},\ref{tRkappa-W}) we obtain
\beq
\tR_{\kappa}(0,t) = t ,
\label{R-kappa-k=0}
\eeq
which agrees with the linear regime prediction associated with Eq.(\ref{kappaL}), i.e.
 the displacement field follows the linear theory at large enough scale.

It is interesting to make the connection between the exact result (\ref{R-kappa-k=0}), which
does not rely on perturbative arguments, and the comparison between Eulerian and Lagrangian propagators. The Galilean transformation (\ref{Galilean}) merely states how particles are
transported over a distance $\vv_0 t$ by long-wavelength modes of the velocity field.
In the Eulerian framework, this "sweeping effect" leads to the Gaussian decay (\ref{R-wnl})
after we take the statistical average over these long-wavelength modes. In contrast, in the
Lagrangian framework the analysis above explicitly shows that this Galilean law 
only sets the normalization at $k=0$ of the propagator, and its values at
$k>0$ probe how trajectories depend on the curvature and higher order derivatives of the
velocity potential, associated with higher orders of the expansion over $k$ of the expression
(\ref{W-def}).
Moreover, in the Lagrangian framework the Galilean transformation
(\ref{Galilean}) appears through relative displacements of well-separated
regions instead of absolute displacements\footnote{As noticed in Sect.~\ref{Lagrangian-prop}, the divergence $\kappa$ is not sensitive to
uniform translations of the system, hence it may seem puzzling that the normalization
(\ref{R-kappa-k=0}) involves the Galilean transformation (\ref{Galilean}). However, this can
be understood as follows. The large-scale limit (\ref{R-kappa-k=0}) and the linear behavior
(\ref{kappaL}) are direct consequences of the requirement that two far-away regions $\vx_1$ and
$\vx_2$, with initial velocities $\vu_{0;1}$ and $\vu_{0;2}$, see their relative distance evolve
as $\vx_2-\vx_1 = t \, (\vu_{0;2}-\vu_{0,1})$. This sets the amplification factor $t$ in Eq.(\ref{kappaL}). 
This configuration considers that regions $1$ and $2$ have constant velocities on some neighborhood and neglects higher order effects, so that they move independently according
to the Galilean law (\ref{Galilean}).}.

\subsubsection{Scaling laws}
\label{scalings}

In arbitrary dimensions computing the averages (\ref{Gkappapsi-nm-0}) or (\ref{W-def})
can lead to lengthy expressions. However, before we explicitly consider the 1D and 2D 
cases, we give some general scaling arguments.
We first need to evaluate the variation of the halo position $\vx$ for infinitesimal variation
of the potential at a summit position $\vq_i$.
By construction, in the $(d+1)$ space $\{\vq,\varphi\}$ the convex hull $\varphi(\vq)$
of Eq.(\ref{phi-convex}) is an hyperplane over the region $\cV$, defined by the $(d+1)$ points $\{\vq_i,\varphi_{L;i}\}$, and its equation reads as
\beq
\det \left( \bea{cccc} \vq_1-\vq_0 & .. & \vq_d-\vq_0 & \vq-\vq_0 \\
\varphi_{L;1}-\varphi_{L;0} & .. & \varphi_{L;d}-\varphi_{L;0} & \varphi-\varphi_{L;0} \ea
\right) = 0 .
\label{hyperplane}
\eeq
Thus, over this cell $\varphi(\vq)$ takes the form
\beq
\varphi(\vq) = \varphi_{L;0} + (\nabla\varphi_L) \cdot (\vq-\vq_{0}) ,
\label{hyperplane-1}
\eeq
where we note $(\nabla\varphi_L)$ a constant vector that is linear over the potentials
$\{\varphi_{L;0},..,\varphi_{L;d}\}$ and that scales as the ratio $(\delta\varphi_L) /\ell$,
where $\ell$ and $(\delta\varphi_L)$ are the typical differences of position and
height between the $(d+1)$ summits ($\ell$ was defined in Eq.(\ref{ell-def})).
From Eq.(\ref{qx-xq}) the position $\vx$ of the associated mass cluster is given
by $\vx=(\nabla\varphi_L)$, and from Eq.(\ref{phiLdef}) we obtain the scaling
\beq
\frac{\pl\vx}{\pl\psi_{0;i}} \sim \frac{t}{\ell}.
\label{Dx-Dpsi0}
\eeq
Substituting into Eq.(\ref{W-def}) we check that the kernel $\tW(k;m)$ is dimensionless
and behaves as
\beq
\tW(k;m) \sim \tW_0(k\ell)  ,
\label{W0-def}
\eeq
in terms of a scaling kernel $\tW_0$.
We have seen in Eq.(\ref{tW-norm}) that $\tW(0;m)=1$, whence $\tW_0(0)=1$.
On the other hand, at high $k$ we can expect the exponential factor in Eq.(\ref{W-def})
to cut large distances beyond $s \sim 1/k$, which gives a power-law decay,
\beq
k \gg \ell^{-1} : \;\; \tW_0(k\ell) \sim (k\ell)^{-d-1} .
\label{W0-highk} 
\eeq
At low masses, numerical simulations and heuristic arguments
\cite{1994A&A...289..325V,2011PhRvD..83d3508V,0755.60104}
suggest that the cluster mass function obeys the  power-law tail
\beq
-3<n<1 , \;\; m \ll \rho_0 L(t)^d : \;\;\; n(m,t) \sim  t^{-d} \,  m^{(n-1)/2} ,
\label{d-nm-lowmass}
\eeq
which has only been proved rigorously in 1D for the Brownian case $n=-2$
\cite{0755.60105,0917.60063,2009JSP...134..589V}
and for the white-noise case $n=0$
\cite{2000JFM...417..323F,0844.35144,2009JSP...137..729V}.
Then, the cutoff (\ref{W0-highk}) gives rise to two different behaviors. For small values of
$n$ the integral (\ref{tRkappa-W}) is dominated by small masses $m$, with $m \propto \ell^d$ and
$\ell \sim 1/k$, because of the cutoff (\ref{W0-highk}) of $\tW_0(k \ell)$, and we obtain the scaling
\beq
n< \frac{2}{d}-1 , \;\; k \gg L(t)^{-1} :  \;\; \tR_{\kappa}(k,t) \sim t \, (kL)^{-d(n+3)/2} ,
\label{Rkapt_highk-d-n1}
\eeq
whereas for large values of $n$ the cutoff (\ref{W0-highk}) is too shallow and the integral is
dominated by large masses (set by the cutoff of the mass function), and lengths $\ell \sim L(t)$
of the order of the typical nonlinear scale (\ref{Lt}), which yields
\beq
n> \frac{2}{d}-1 , \;\; k \gg L(t)^{-1} :  \;\; \tR_{\kappa}(k,t) \sim t \, (kL)^{-d-1} .
\label{Rkapt_highk-d-n2}
\eeq

The scaling laws (\ref{Rkapt_highk-d-n1},\ref{Rkapt_highk-d-n2}) have been derived assuming
that small-mass Lagrangian cells are characterized by a single scale $\ell$, which leads to
Eq.(\ref{W0-def}). 
As we discuss in Sect.~\ref{2Dcase} below, in dimensions greater than one small cells may
be characterized by several lenght scales, which behave as power laws over mass with
different exponents. This would violate the scaling laws
(\ref{Rkapt_highk-d-n1},\ref{Rkapt_highk-d-n2}) and give rise to new exponents.

\subsubsection{1D case}
\label{1Dcase}

In 1D the computations greatly simplify, since cells are mere intervals characterized
by a single scale, their length $\ell$ or equivalently their mass $m=\rho_0 \ell$.
In agreement with Eq.(\ref{hyperplane}), the convex hull $\varphi(q)$ between the
boundary points $q_0$ and $q_1$ is the straight line
\beq
\varphi(q)= \varphi_{L;0}+\frac{q-q_0}{q_1-q_0} \, (\varphi_{L;1}-\varphi_{L;0}) ,
\eeq
the Eulerian position $x$ of the cluster (shock) is
\beq
x= \frac{\varphi_{L;1}-\varphi_{L;0}}{q_1-q_0} ,
\eeq
and its derivative reads as
\beq
\frac{\pl x}{\pl \psi_{0;0}} = \frac{t}{q_1-q_0} .
\eeq
The integral over $\vs$ in Eq.(\ref{Gkappapsi-nm-0}) is now the discrete sum over the
two boundary points, $q_0$ and $q_1$, and taking the average over the two orientations
of the interval with respect to $q_0=0$, that is, either $q_1=\ell$ or $q_1=-\ell$,
we obtain
\beqa
\pR_{\kappa\psi}(\Delta q,t) & = & \int_0^{\infty} \dd m \, n(m) \, \frac{t}{\ell} 
\left[ \delta_D(\Delta q +\ell) + \delta_D(\Delta q -\ell) \right. \nonumber \\
&& \left. - 2 \delta_D(\Delta q) \right] .
\label{R-kappa-psi-1d}
\eeqa
This is the result that was already obtained in \cite{2010PhRvD..81d3516B},
which yields in Fourier space
\beq
\tW(k;m) = \frac{2}{k^2\ell^2} \, [1-\cos(k\ell)] ,
\eeq
and
\beq
\tR_{\kappa}(k,t) = \int_0^{\infty} \dd m \, n(m) \, \frac{2t}{k^2\ell} \, \left[1- \cos(k\ell) \right] .
\label{tR-kappa-1d}
\eeq
This also means that we recover the high-$k$ decay (\ref{W0-highk}) and
the high-$k$ power-law (\ref{Rkapt_highk-d-n1}) is indeed realized for
all values of $n$ in the range $-3<n<1$ that we consider in this paper, because for high $k$ the
integral (\ref{tR-kappa-1d}) is always governed by the region $\ell \sim 1/k$.
Thus, we recover the results (\ref{W0-highk}) and (\ref{Rkapt_highk-d-n1}) that were obtained
in Sect.~\ref{scalings} from simple dimensional and scaling arguments.
This is not surprising since in 1D Lagrangian cells are characterized by a single scale.

\subsubsection{2D case}
\label{2Dcase}

In 2D the computation follows the same route, the Lagrangian cells $\cV$ being now triangular.
Let us consider a triangle of summits $\{S_{1},S_{2},S_{3}\}$, inner angles $\{\alpha_{1},\alpha_{2},\alpha_{3}\}$, and
opposite-side lengths $\{s_{1},s_{2},s_{3}\}$. We can also choose the triangle to be positively oriented,
that is, $\lvec{S_{1}S_{2}}\times\lvec{S_{1}S_{3}}=s_{2}s_{3}\sin(\alpha_{1})\ve_3$, where $\{\ve_1,\ve_2,\ve_3\}$ is a 3D
right-handed coordinate system and the triangle is in the plane $\{\ve_1,\ve_2\}$.
Then, from Eq.(\ref{hyperplane}) we obain the equation $\varphi(\vq)$ of the planar facet
that goes through the three summits, and from Eq.(\ref{qx-xq}) the position of the mass
cluster is $\vx=\pl\varphi/\pl\vq$. Taking the derivative with respect to $\psi_{0;S_{1}}$,
from Eq.(\ref{phiLdef}), we obtain
\beq
\frac{\pl\vx}{\pl\psi_{0;S_{1}}} = \frac{t}{2\ell^2} \; \lvec{S_{2}S_{3}}\times\ve_3 ,
\eeq
where $\ell^2$ is now the triangle area, and the scalar product with the outer normal
$\vn_{\rm out}$ along each triangle side reads as
\beqa
(S_{1}S_{2}) & : & \vn_{\rm out}^{(S_{1}S_{2})} \cdot \frac{\pl\vx}{\pl\psi_{0;S_{1}}} = \frac{-t s_{1}}{2\ell^2} \cos\alpha_{2} , \\
(S_{2}S_{3}) & : & \vn_{\rm out}^{(S_{2}S_{3})} \cdot \frac{\pl\vx}{\pl\psi_{0;S_{1}}} = \frac{t s_{1}}{2\ell^2} , \\
(S_{3}S_{1}) & : & \vn_{\rm out}^{(S_{3}S_{1})} \cdot \frac{\pl\vx}{\pl\psi_{0;S_{1}}} = \frac{-t s_{1}}{2\ell^2} \cos\alpha_{3} .
\eeqa
Then, the kernel $\tW$ reads from Eq.(\ref{W-def}) as
\beqa
\tW(k;m) & = & \biggl\lag \frac{3s_{1}}{2k^2\ell^4} \int_0^1 \dd u \; [ s_{3} \, \cos\alpha_{2} \, e^{-\ii u \vk\cdot\vq_{S_{2}}}  \nonumber \\
&& \hspace{-2cm} - s_{1} \, e^{-\ii\vk\cdot[\vq_{S_{2}}+u(\vq_{S_{3}}-\vq_{S_{2}})]} + s_{2} \, \cos\alpha_3 \, e^{-\ii u \vk\cdot\vq_{S_{3}}} ] \biggl\rag_{\!m} 
\eeqa
where we chose the origin of coordinates as $\vq_{S_{1}}=0$.
Taking the average over the global direction of the triangle (which is equivalent to averaging over
the direction of $\vk$), we obtain
\beqa
\tW(k;m) \!\! & \!=\! & \biggl\lag \frac{3s_{1}}{2k^2\ell^4} \!\int_0^1\!\! \dd u \bigg[ s_{3} \, \cos\alpha_{2} \, J_0(uks_{3})
 \nonumber \\
&&+ s_{2} \, \cos\alpha_{3} \, J_0(uks_{2}) -s_{1} \nonumber \\
&& \hspace{-2cm}\times J_0(k\sqrt{{s_{2}}^2u^2\!+\!{s_{3}}^2(1\!-\!u)^2\!+\!2s_{2}s_{3} u(1\!-\!u) \cos\alpha_{1}}) \bigg] \biggl\rag_{\!m} .
\label{tW-m-J0}
\eeqa
We can be here a bit more explicit in the ensemble average calculations. Let us define the angle distribution $P_{\rm tri.}(\alpha_{1},\alpha_{2};m)\dd\alpha_{1}\dd\alpha_{2}$ for a give mass (i.e. area) with the relations (with $\alpha_{3}=\pi-\alpha_{1}-\alpha_{2}$),
\begin{equation}
\ell^2=\frac{1}{2}{s_2}s_{3}\sin\alpha_{1},
\end{equation}
\begin{equation}
s_{3}^2=2\ell^2\left(\cot\alpha_{1}+\cot\alpha_{2}\right)
\end{equation}
and those obtained by circular permutations. As a result the expression (\ref{tW-m-J0}) can be rewritten
\beqa
\tW(k;m) & = & 
\frac{3}{k^2\ell^2} \!\int_0^1\!\! \dd u\, \int P_{\rm tri.}(\alpha_{1},\alpha_{2};m) \dd\alpha_{1}\dd\alpha_{2}
 \nonumber \\
&&  \hspace{-2cm}\bigg[ \cot\alpha_{2} \, J_0(uks_{3})
+ \cot\alpha_{3} \, J_0(uks_{2}) -\left(\cot\alpha_{2}+\cot\alpha_{3}\right) \nonumber \\
&& \hspace{-2cm}\times J_0(k\sqrt{{s_{2}}^2u^2\!+\!{s_{3}}^2(1\!-\!u)^2\!+\!2s_{2}s_{3} u(1\!-\!u) \cos\alpha_{1}}) \bigg] .
\label{tW-m-Ptr}
\eeqa
Note that expanding the Bessel functions up to order $k^2$ and using standard relations between
the angles, sides and area of triangles, the property (\ref{tW-norm}) can be easily recovered.

\begin{figure}
\begin{center}
\epsfxsize=7. cm  {\epsfbox{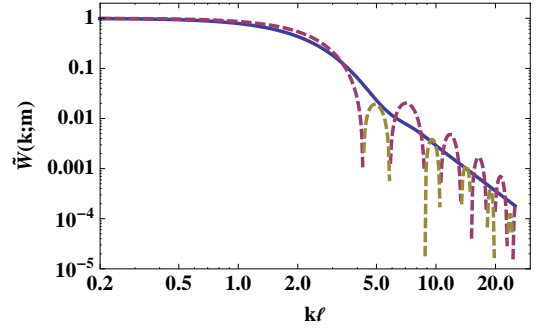}}
\end{center}
\caption{The resulting kernel function function $\tW_{\kappa}(k;m)$ for equilateral triangles (dashed line) and for the angle distribution of a Delaunay triangulation of a Poisson point process (solid line). }
\label{kPropagator2D}
\end{figure}

The actual distribution $P_{\rm tri.}$ to use actually depends on the details
of the Lagrangian-space tessellation, which in turn depend on the index $n$ of the initial conditions.
For instance, if all triangles were equilateral, we would have
\beq
\tW_{\rm eq}(k;m) = \frac{8}{k^2s^2} \int_0^1 \dd u \, [ J_0(uks) - J_0(ks\sqrt{u^2-u+1})]
\eeq
with $s^2=(4m)/(\sqrt{3}\rho_0)$.
Another interesting, and probably more realistic, case is provided by the angle distribution obtained in the Delaunay 
triangulation of a Poisson point process, for which we have,~\cite{Miles197085,1987A&A...184...16I},
\begin{eqnarray}
P_{\rm tri.}(\alpha_{1},\alpha_{2})\dd\alpha_{1}\dd\alpha_{2}&=&\nonumber\\
&&\hspace{-3cm}
\frac{8}{3\pi}\sin(\alpha_{1})\sin(\alpha_{2})\sin(\alpha_{1}+\alpha_{2})\dd\alpha_{1}\dd\alpha_{2}
\end{eqnarray}
with
\begin{equation}
0<\alpha_{1}<\pi,\ 0<\alpha_{2}<\pi,\ 0<\alpha_{1}+\alpha_{2}<\pi,
\end{equation}
independently of the area $\ell^2$ of the considered triangles.
The resulting angle integration cannot be performed  explicitly. We instead propose
a fit to the resulting dimensionless kernel,
\begin{eqnarray}
\tW(k;m) & = & 0.749428 \, e^{-0.206885 (k\ell)^2} \nonumber \\
&& \hspace{-1cm} 
+\frac{0.250572}{\sqrt{0.00707579 (k\ell)^6+1.00806 (k\ell)^2+1}}.
\label{tW-fit}
\end{eqnarray}
The resulting shapes of the kernel are shown on Fig. \ref{kPropagator2D}.

These forms illustrate the high $k$ asymptotic forms of the propagators.
At fixed triangle area and shape, the integration over the Bessel functions in (\ref{tW-m-J0})
leads to a high-$k$ power-law decay of the form $\tW(k;m) \sim (k\ell)^{-3}$, in agreement with
Eq.(\ref{W0-highk}). This gives the $n$-dependent decay $\tR_{\kappa}(k,t) \sim t (kL)^{-n-3}$
of Eq.(\ref{Rkapt_highk-d-n1}) for $n<0$, and the constant-slope decay
$\tR_{\kappa}(k,t) \sim t (kL)^{-3}$ of Eq.(\ref{Rkapt_highk-d-n2}) for $n>0$.

However, these exponents could be modified if the distribution of triangle
shapes, unlike the two cases considered above, 
gives significant weight to configurations where the three sides have very different
lengths. For instance, in the extreme case where triangles are increasingly ``squeezed'', with
two sides that remain of order $L(t)$ and a third side $\epsilon$ that scales with mass
as $m \sim \epsilon L$, the kernel (\ref{tW-m-J0}) is dominated by the case where the summit $S_1$
is at one end of the small side and it behaves at high $k$ (and small mass) as
\beq
m \sim \epsilon L , \;\; \tW_{\rm squeezed}(k;m) \sim k^{-3} \epsilon^{-2} L^{-1} .
\eeq
This now gives over the full range $-3<n<1$ the high-$k$ decay
\beq
k \gg L(t)^{-1} : \;\; \tR_{\kappa,{\rm squeezed}}(k,t) \sim t \, (kL)^{-3(n+3)/4} ,
\label{R-squeezed}
\eeq
which is shallower than the scalings (\ref{Rkapt_highk-d-n1})-(\ref{Rkapt_highk-d-n2}).
This shows how the results of Sect.~\ref{scalings}, which rely on simple scaling arguments and
dimensional analysis, can be violated in complex cases where Lagrangian cells of a given mass
involve several scales with different characteristic exponents.

We can expect the scalings (\ref{R-squeezed}) and (\ref{Rkapt_highk-d-n1})-(\ref{Rkapt_highk-d-n2})
to bracket the high-$k$ exponents that can be reached for arbitrary triangle distributions,
depending on scaling or on the fraction of ``squeezed'' triangles as a function of mass.
This holds for  mass functions that satisfy the low-mass power laws (\ref{d-nm-lowmass}), but
for more general cases we could extend this analysis to arbitrary low-mass behaviors of the
mass functions.
In any case, these results show that the Lagrangian propagator is a sensitive probe of the structure
of the matter distribution. Moreover, in dimensions 2 and higher, it involves both the cluster
mass function and key properties of the shape of Lagrangian-space tessellations, that is,
of the shape of the initial regions that eventually end up in small-mass clusters.

Another significant difference with the Eulerian propagators discussed in
Sect.~\ref{Eulerian-propagators}, which vanished for $-3<n<-1$ because of infrared
divergences of the initial velocity field, is that the Lagrangian propagators are finite and well-defined over the full range $-3<n<1$.

\section{Numerical simulations}
\label{Numerical}


We now describe the results that we obtain from a numerical study, in both the
1D and 2D cases. This allows us to check the exact results presented in
Sect.~\ref{Analytical-results} and our predictions (for $d= 2$),
which we think are illustrative of what is happening at higher dimension.
The details of our numerical simulations, and of the new efficient algorithms that
we use, are given in \cite{2011PhRvD..83d3508V}. There,  the statistical properties of the matter
distributions built by the geometrical adhesion model defined in
Sect.~\ref{eq-motion}, for the same Gaussian scale-invariant initial conditions
given in Sect.~\ref{Initial-conditions}, are described in details.
In particular, it is shown that the density probability distributions and the
mass functions exhibit qualitatively the same properties as those observed for
3D gravitational clustering (see also
\cite{1990MNRAS.242..200K,1994A&A...289..325V}).

Evolving from the initial Gaussian, that remains relevant on large scales and
at early times, the density probability distribution gradually broadens and
builds an intermediate power-law regime, in-between rare voids and 
rare overdensities. On the other hand, the shock mass function shows
a low-mass power tail and a high-mass exponential-like falloff, and obeys
up to a good accuracy the Press-Schechter-like scaling in terms of the
reduced variable $\nu=\delta_L/\sigma(M)$ \cite{1974ApJ...187..425P}.
However, while in 1D the scaling mass function $f(\nu)$ agrees quite
well with the original Press-Schechter ansatz (which actually happens to
be exact in the case $n=-2$ \cite{2009JSP...134..589V}), in 2D it is significantly
different and it shows a $\nu^2$ low-mass tail instead of the 
Press-Schechter prediction $\propto \nu$.

These studies have shown that the adhesion model shares many properties
with the gravitational dynamics, which motivates a further investigation
with respect to the response functions and correlators, which are critical
ingredients in analytical approaches to gravitational clustering. 

The advantages for using the adhesion model are discussed in \citep{2011PhRvD..83d3508V} 
where one can further find the description
of the algorithms we use. Let us remind that our method allows us to get the mass distribution without 
introducing further discretization, either in time or in space.
We can also cover a greater range of masses and scales as compared with
usual gravitational dynamics (especially as we consider lower-dimensional
systems, either 1D or 2D), which gives us a better control of asymptotic regimes.
For numerical purposes, the power-law initial conditions (\ref{PdeltaL})
also improve the statistics since by using the self-similarity (\ref{selfsimilar}) 
we can rescale coordinates and propagators to obtain time-independent functions, so that
we can take the mean over all output times when we compute statistical averages.

More precisely, it is convenient to introduce the dimensionless scaling variables
\beqa
\vQ= \frac{\vq}{L(t)} , \;\;\; \vX= \frac{\vx}{L(t)} , \;\;\; \vK= L(t) \vk , \nonumber \\
\vU= \frac{t\vu}{L(t)} , \;\;\; M= \frac{m}{\rho_0 L(t)^d} .
\label{QXU}
\eeqa
Then, equal-time statistical quantities (such as correlations or probability
distributions) written in terms of these variables no longer depend on time and the
scale $X=1$ is the characteristic scale of the system, associated with the transition
from the linear to nonlinear regime.
In terms of the propagators, this gives for instance
\beq
\pR_{\delta}(x,t) = \frac{t}{L(t)^d} \, \pR_{\delta}(X) , \;\;\; \tG_{\delta}(k,t) = t \,
\tG_{\delta}(K) .
\label{R-selfsim}
\eeq
In particular in the linear regime we have,
\begin{equation}
\pR^{{\rm lin}}_{\delta}(\vX)=\delta_{D}(\vX), \ \ \tG^{{\rm lin}}_{\delta}(K)=1.\label{RL-selfsim}
\end{equation}
The Lagrangian propagator $\tR_{\kappa}$ obeys the same relations.

As in \cite{2011PhRvD..83d3508V}, we consider the cases $n=0.5, 0, -0.5, -1, -1.5, -2$, and
$-2.5$ to cover the range $-3<n<1$ that we study in this article, where
the self-similarity (\ref{selfsimilar}) holds. This is also the range of interest
for cosmological purposes. In particular, the definition of the slope $n$
in Eq.(\ref{PdeltaL}) is such that, whatever the dimension $d$ that we
consider, it corresponds to the usual slope $n$ that appears in papers on
3D gravitational clustering in cosmology (thus, with this choice the scaling law
(\ref{selfsimilar}) does not depend on $d$).

We focus on the Fourier-space propagators in the main text because they
are the quantities that appear in the resummation schemes used in cosmology,
However, we briefly describe in App.~\ref{real-space} our results in real space,
for the response function $\pR_{\psi}(\vX)$. This is of interest by itself because
for the Burgers dynamics this is also the one-point velocity probability distribution,
as seen in Eq.(\ref{Rpsi-pxu}). Moreover, in the nonlinear regime real-space
methods may prove as useful as the Fourier-space methods that have been
developed so far.

\subsection{Eulerian propagators}
\label{EulerianPropagators}

\subsubsection{Numerical estimates of propagators}
\label{NumericalEstimates}

To compute the Fourier-space response function we use the definition (\ref{tR-def}),
as the mean functional derivative of the fields in Fourier space.
Thus, taking as a reference an initial condition $\psi_0(\vx)$, i.e. $\tpsi_0(\vk)$
in Fourier space,
we obtain the nonlinear field $\psi(\vx,t)$, whence $\psi(\vk,t)$, at time $t$.
Then, we perturb the initial condition at a given wavenumber $\vk_0$.
Since all fields are real, this means that we consider the change
$\psi_0(\vx)\rightarrow \psi_0(\vx)+\Delta\psi_0(\vx)$ with
$\Delta\psi_0(\vx)= \epsilon \, 2 \cos(\vk_0\cdot\vx)$, with a small amplitude $\epsilon$.
Then, we compute the nonlinear field $\psi+\Delta\psi$ at time $t$ generated
by this new initial condition, we take its Fourier transform, and we obtain the
functional derivative as $\cD\tpsi/\cD\tpsi_0= \Delta \tpsi(\vk_0)/\epsilon$.
Next, to estimate the response function $\tR_{\psi}(k_0,t)$ as in Eq.(\ref{tR-def}), with the
statistical averaging $\langle .. \rangle$, we repeat the operation above for many
reference initial conditions $\psi_0$, with the Gaussian statistics
of Sect.~\ref{Initial-conditions}, and the mean over all these simulations
gives our final estimate of $\tG_{\psi}(k_0,t)$.
In 2D we perturb the initial condition along the two axis, $\vk_0=k_0 \ve_1$ and
next $\vk_0=k_0 \ve_2$, which provides two measures of the functional derivative.
Finally, using the self-similarity (\ref{selfsimilar}) we rescale the results obtained at different
output times and we make a final averaging to increase the statistical accuracy,
to obtain the dimensionless propagators as in Eq.(\ref{R-selfsim}).
A similar procedure also provides the propagator $\tR_{\delta}$.
Note finally that for the 1D case and for the reduced variable $K$, we have
\begin{equation}
\tG^{1\rm D}_{\delta}(K) =\tG^{1\rm D}_{\psi}(K) , 
\end{equation}
but this identity is not true for higher dimensions, as shown in Sect.~\ref{Eulerian-propagators}.

Instead of computing the response functions $\tR(K)$ through functional derivatives
as in Eq.(\ref{tR-def}), it is possible to derive these propagators through
cross-correlations as in Eq.(\ref{tP-def}). We also considered this alternative
method to check our numerical algorithms. This also provides a confirmation of
the general identities (\ref{S-def}) and (\ref{R-funct-2}) into the shell-crossing
regime. We briefly show a comparison of these two procedures in
App.~\ref{Cross-correlations} for the 1D case (we obtain similar but somewhat
more noisy results in 2D).

\subsubsection{One-dimensional dynamics}
\label{One-dimensionalEulerian}

\begin{figure}
\begin{center}
\epsfxsize=8.5 cm \epsfysize=6 cm {\epsfbox{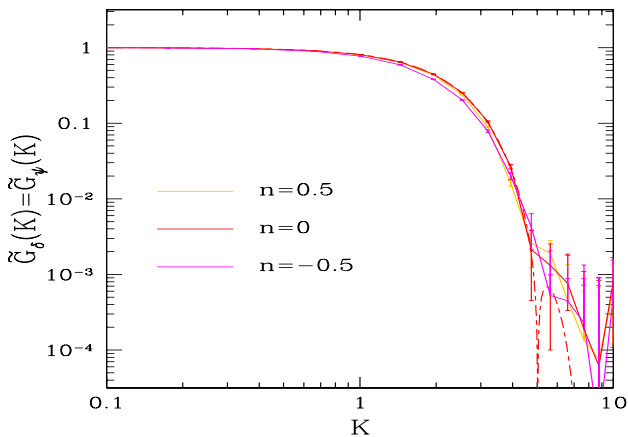}}
\end{center}
\caption{The 1D Fourier-space propagator $\tG_{\delta}(K)=\tG_{\psi}(K)$
from numerical simulations. We show the cases $n=0.5, 0$, and 
$-0.5$, as a function of the reduced wavenumber $K$. For $n=0$ we also
plot the exact analytical result (dot-dashed line) from \cite{2010PhRvD..81d3516B}.}
\label{figRpsi-1d}
\end{figure}

We show in Fig.~\ref{figRpsi-1d} the Fourier-space Eulerian response function
$\tG_{\psi}(K)=\tG_{\delta}(K)$, in terms of the reduced wavenumber $K$
of Eq.(\ref{QXU}) for the 1D case. From Eqs.(\ref{Rpsi-pxu}) and (\ref{Rdelta-pxu}),
we have in real space
\beq
\pR_{\delta}(X) = \pR_{\psi}(X) = P(U) , \;\; \mbox{with} \;\; U=X ,
\label{Rpsi-delta-PU}
\eeq
where $P(U)$ is the one-point probability distribution of the reduced velocity $U$,
so that  $\tR_{\delta}(K)$ and $\tR_{\psi}(K)$ are also the Fourier transforms of the
velocity distribution.
Since for $-3<n<-1$ the response functions are not well defined because of the
divergent ``sweeping effect'', we only show our results for the cases
$n=0.5, 0$, and $-0.5$. For the case $n=0$ we also plot the exact result,
given by Eq.(65) and Figs.~1 and 2 in \cite{2010PhRvD..81d3516B}.
We can check that our numerical result agrees quite well with this analytical
result and its strong exponential-like decay $\sim e^{-K^{3/2}}$.
However, the numerical errorbars are too large to distinguish the oscillatory
behavior (i.e., changes of sign) in the far tail, where $\tR < 10^{-3}$.
Our numerical results show that this response function obeys a similar
exponential-like decay at high $K$
for $n=0.5$ and $n=-0.5$. A priori the falloff can depend on $n$,
as $\tG \sim e^{-K^{\alpha}}$ with some exponent $\alpha(n)$.
The range of scales available in Fig.~\ref{figRpsi-1d} is too small to
obtain a precise measure of $\alpha(n)$, but it suggests that $\alpha$ does not
vary too much over $-0.5\leq n \leq 0.5$.

For $K\rightarrow 0$ we recover $\tR_{\psi}\rightarrow 1$, in agreement with
Eq.(\ref{R-psi-k0}) and with the normalization to unity of $P(U)$ in
Eq.(\ref{Rpsi-delta-PU}).

\subsubsection{Two-dimensional dynamics}
\label{Eulerian-2d}

\begin{figure}
\begin{center}
\epsfxsize=8.5 cm \epsfysize=6 cm {\epsfbox{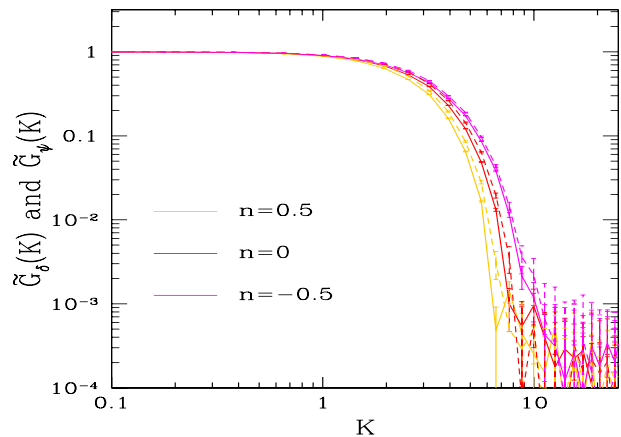}}
\end{center}
\caption{The 2D Fourier-space propagators $\tR_{\delta}(K)$ (solid line) and
$\tR_{\psi}(K)$ (dashed line).}
\label{figGdeltaGpsi-2d}
\end{figure}

Let us now consider the 2D case.
We show in Fig.~\ref{figGdeltaGpsi-2d} the 2D Fourier-space propagators
$\tR_{\delta}(K)$ and $\tR_{\psi}(K)$. 
As in the 1D case of Fig.~\ref{figRpsi-1d}, we can clearly see the steep exponential 
cutoff, but with a slightly stronger dependence on the exponent $n$ of the initial conditions. We also recover the exact low-$K$ limit $\tR_{\psi}(0)=1$.

As explained in Sect.~\ref{Eulerian-propagators}, in dimension greater than one
the density and velocity potential propagators are no longer
identical, because of the nonlinear relationship (\ref{delta-psi-d}), but they
are all expected to be governed by the ``sweeping effect''.
This is confirmed by our numerical results, since we clearly see in
Fig.~\ref{figGdeltaGpsi-2d} that $\tR_{\delta}(K)$ shows the same
exponential-like cutoff at high $K$. In fact, in 2D the density propagator is
very close to its velocity-potential counterpart,
although there are hints of a slightly steeper falloff for the density response function.

\subsection{Lagrangian propagators}
\label{LagrangianPropagators}

As for the Eulerian propagators studied in Sect.~\ref{EulerianPropagators},
we measure the response function $\tR_{\kappa}(K)$ from its definition
(\ref{tR-def-Lag2}),
by estimating the functional derivative from the difference between two very
close initial conditions.
In contrast to the Eulerian propagators, the Lagrangian propagators
are well defined over the full range $-3<n<1$, hence we plot
the cases $n=0.5, 0, -0.5, -1, -1.5, -2$, and $-2.5$.

\subsubsection{One-dimensional dynamics}
\label{Lagrangian-1d}

\begin{figure}
\begin{center}
\epsfxsize=8.5 cm \epsfysize=6 cm {\epsfbox{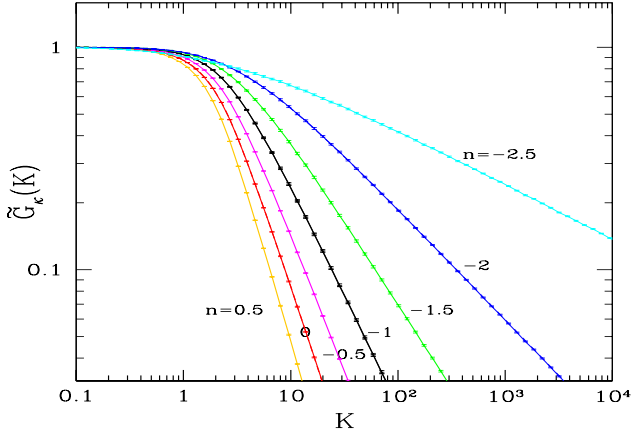}}
\end{center}
\caption{The Fourier-space response function $\tR_{\kappa}(K)$ (solid line)
from 1D numerical simulations. For $n=0$ and $n=-2$ we also plot the exact
analytical result (dot-dashed line) from \cite{2010PhRvD..81d3516B}.}
\label{figRkappa-1d}
\end{figure}

\begin{figure}
\begin{center}
\epsfxsize=8.5 cm \epsfysize=6 cm {\epsfbox{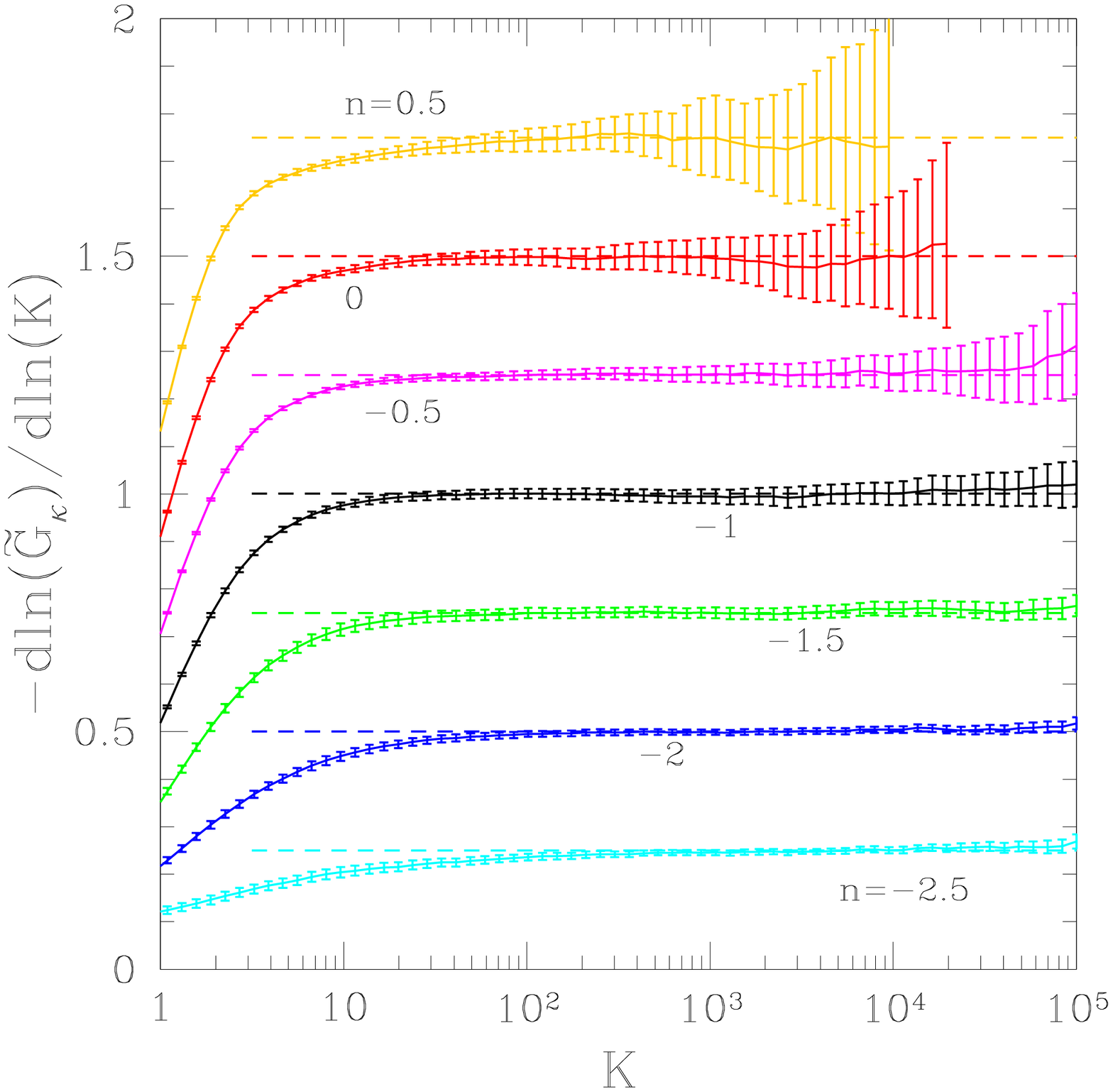}}
\end{center}
\caption{The logarithmic derivative, $-\dd\ln\tR_{\kappa}/\dd\ln K$, of
the Lagrangian propagator, from 1D numerical simulations (solid line), and its theoretical asymptotic limit $(n+3)/2$ (dashed line).}
\label{figdlRkappadlk-1d}
\end{figure}

We show in Fig.~\ref{figRkappa-1d} the Fourier-space Lagrangian propagator 
$\tR_{\kappa}(K)$.
For the two cases $n=0$ and $n=-2$ we obtain a very good agreement between
our numerical results and the exact analytical results derived in 
\cite{2010PhRvD..81d3516B} (the numerical and analytical curves cannot even
be distinguished in this figure).
We recover the exact low-$K$ limit $\tR(0)=1$, which corresponds to 
Eq.(\ref{R-kappa-k=0}). 
As explained in Sect.~\ref{sum-rules}, this is related to the relative motions
of well-separated regions by long-wavelength modes of the velocity field, and it
applies to the full range $-3<n<1$.

Figure~\ref{figRkappa-1d} clearly shows a power-law tail at high $K$,
contrary to the exponential-like cutoff seen in Fig.~\ref{figRpsi-1d}, with a
slope that depends on $n$. To clearly see this power-law tail we show in
Fig.~\ref{figdlRkappadlk-1d} the logarithmic derivative
$-\dd\ln\tR_{\kappa}/\dd\ln K$, as well as its asymptotic theoretical prediction
from Eq.(\ref{tR-kappa-1d}), which agrees with Eq.(\ref{Rkapt_highk-d-n1})
and reads as
\beq
d=1 , \;\; -3<n<1 , \;\; K \rightarrow \infty : \;\; 
\frac{\dd\ln\tR_{\kappa}}{\dd\ln K} \rightarrow - \frac{n+3}{2} .
\label{Rkappa-high-k-d-1}
\eeq
We can check that our numerical results are consistent with this analytical
result and we clearly see the convergence to the high-$K$ slope 
(\ref{Rkappa-high-k-d-1}).

\subsubsection{Two-dimensional dynamics}
\label{Lagrangian-2d}

\begin{figure}
\begin{center}
\epsfxsize=8.5 cm \epsfysize=6 cm {\epsfbox{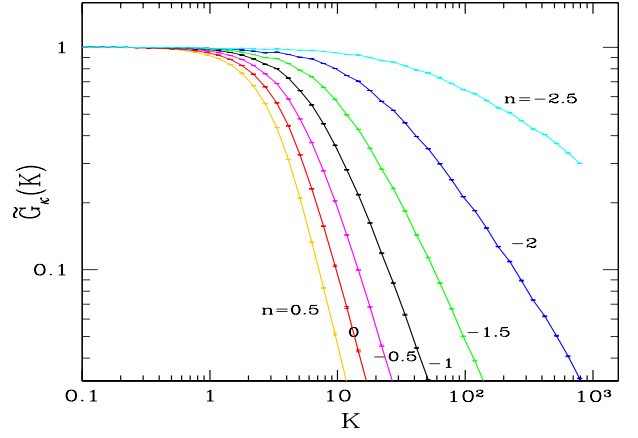}}
\end{center}
\caption{The 2D Fourier-space propagator $\tR{\kappa}(K)$ (solid line)
from numerical simulations.}
\label{figRkappa-2d}
\end{figure}

\begin{figure}
\begin{center}
\epsfxsize=8.5 cm \epsfysize=6 cm {\epsfbox{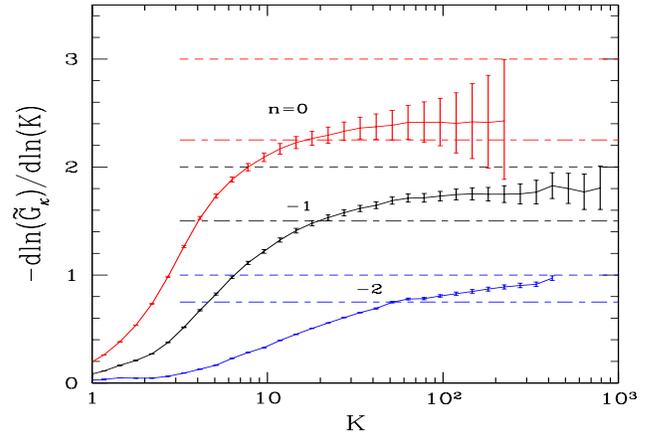}}
\end{center}
\caption{The logarithmic derivative, $-\dd\ln\tR_{\kappa}/\dd\ln K$, of
the Lagrangian propagator, from 2D numerical simulations (solid line). We also 
show the two theoretical predictions of Eq.(\ref{Rkappa-high-k-d-2-equil})
(upper dashed line) and Eq.(\ref{Rkappa-high-k-d-2-squeezed}) 
(lower dot-dashed line).}
\label{figdlRkappadlk-2d}
\end{figure}

We show in Fig.~\ref{figRkappa-2d} the Fourier-space Lagrangian response
function $\tR_{\kappa}(K)$.
As in the 1D case plotted in Fig.~\ref{figRkappa-1d}, we recover the low-$K$
limit $\tR_{\kappa}(0)=1$. We again obtain a clear power-law tail at high $K$,
but the convergence to this asymptotic regime is somewhat slower than in
1D, especially for low values of $n$.

To estimate the high-$K$ exponent we plot in Fig.~\ref{figdlRkappadlk-2d}
the logarithmic derivative, $-\dd\ln\tR_{\kappa}/\dd\ln K$.
As explained in Sect.~\ref{2Dcase}, depending on the shape of the
Lagrangian-space tessellation, we have two different predictions for the
extreme cases of small-mass triangles characterized by a single scale
$\ell \sim m^{1/2}$, that is, triangles of approximately ``equilateral'' shape,
\beq
\mbox{``equil.''} , \;\; K \rightarrow \infty : \;\; 
\frac{\dd\ln\tR_{\kappa}}{\dd\ln K} \rightarrow - \min(n+3,3) ,
\label{Rkappa-high-k-d-2-equil}
\eeq
which follows from Eqs.(\ref{Rkapt_highk-d-n1}), (\ref{Rkapt_highk-d-n2}),
and for ``squeezed'' triangles, where two sides remain on the order of
the nonlinear scale $L(t)$ whereas the third side decreases as $\ell_3 \sim m$,
\beq
\mbox{``squeezed''} , \;\; K \rightarrow \infty : \;\; 
\frac{\dd\ln\tR_{\kappa}}{\dd\ln K} \rightarrow - \frac{3(n+3)}{4} ,
\label{Rkappa-high-k-d-2-squeezed}
\eeq
from Eq.(\ref{R-squeezed}).
We clearly see in Fig.~\ref{figdlRkappadlk-2d} the convergence towards
a constant logarithmic slope,  although for low $n$ the convergence is quite slow
and is not complete yet on this range. As expected the high-$K$ exponent
is between the two limiting values (\ref{Rkappa-high-k-d-2-equil}) and
(\ref{Rkappa-high-k-d-2-squeezed}). Unfortunately, it also seems that it is
different from these two values, which means that the triangulation is neither
dominated by ``equilateral'' (i.e. ``single-scale'') triangles nor by ``maximally squeezed'' triangles. The Lagrangian tessellation is probably more complex
than these two simple cases and Fig.~\ref{figdlRkappadlk-2d} suggests that
there is a broad distribution of shapes for low-mass triangles.

This agrees with a qualitative inspection of Fig.~6 in 
\cite{2010PhRvE..82a6311B}, where we showed two examples of Lagrangian
triangulations obtained for $n=0$ and $n=-2$.
There, one can clearly see that the summits of the Lagrangian cells are not 
distributed at random but show a strong clustering. One can see distant groups
of Lagrangian summits, separated by a length of order $L(t)$, with small triangles 
within each group (they would correspond to low-mass ``equilateral'' shapes) and 
very thin triangles that join two summits in one group to a third summit in a 
second group (they would correspond to low-mass ``squeezed'' shapes). 
This suggests that the Lagrangian-space tessellation contains both ``equilateral'' 
and ``squeezed'' configurations. However, it is not clear whether there is a
bimodal or a continuous distribution of triangle shapes.

For completeness, we should point out that numerical results suggest that
the number of mass clusters per unit Lagrangian or Eulerian volume 
is infinite if $-3<n<-1$ and finite if $-1<n<1$, as seen from the low-mass
tail (\ref{d-nm-lowmass}) of the mass function 
\cite{1994A&A...289..325V,2011PhRvD..83d3508V}.
Moreover, it appears that clusters are dense in Eulerian space for $-3<n<-1$ 
while they are isolated for $-1<n<1$. In Lagrangian space the geometry is
somewhat more complex. For $-1<n<1$ there is still a finite number of cells
and summits per unit area, but for $-3<n<-1$ the infinite number of triangles
and summits does not cover the whole plane. This is obvious from the fact that
there are some large triangles associated with finite-mass clusters.
Therefore, the infinitely many summits ``gather'' in some regions of Lagrangian
space, in-between the large triangles associated with massive 
halos\footnote{These Eulerian- and Lagrangian-space behaviors are also
found in the 1D case, where they are rigorously proved for $n=-2$ and $n=0$.}.
Nevertheless, these rather different properties of the Eulerian and
Lagrangian tessellations with $n$ \cite{2010PhRvE..82a6311B} do not seem
to have a strong impact on the low-mass tail nor on the high-$K$
asymptote of the Lagrangian response function
(e.g., there is no clear sign in Fig.~\ref{figdlRkappadlk-2d} that the exponent
switches from (\ref{Rkappa-high-k-d-2-equil}) to
(\ref{Rkappa-high-k-d-2-squeezed}) as $n$ goes to either side of $-1$).

\subsubsection{2D Triangle shape distribution}
\label{Shapes-2d}

\begin{figure}
\begin{center}
\epsfxsize=8.5 cm \epsfysize=6 cm {\epsfbox{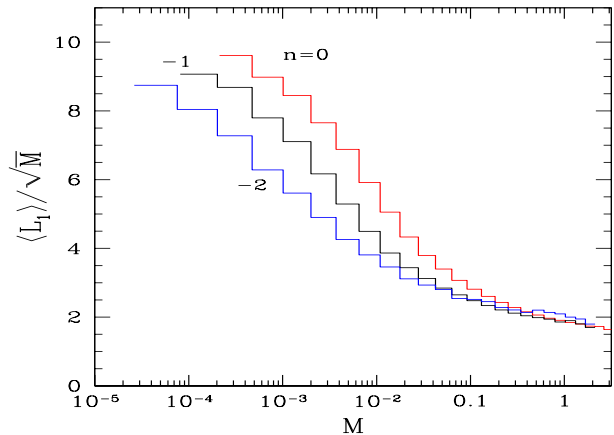}}
\end{center}
\caption{The statistical average $\langle L_1\rangle$ of the longest
triangle side $L_1$. We plot this quantity for a series of triangle mass bins 
and we divide $\langle L_1\rangle$ by the factor $\sqrt{M}$, where $M$ is the
mean mass of each bin.}
\label{figl1-2d}
\end{figure}

\begin{figure}
\begin{center}
\epsfxsize=8.5 cm \epsfysize=6 cm {\epsfbox{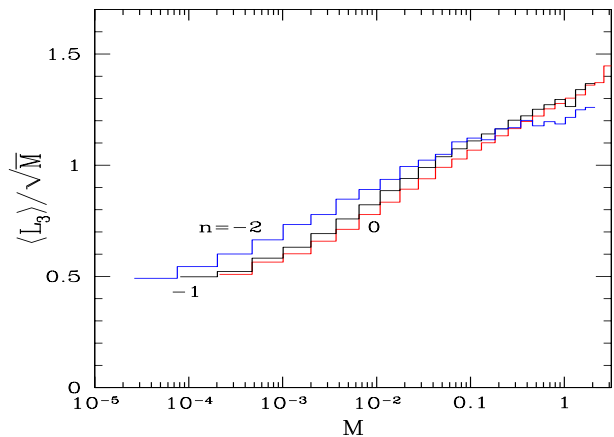}}
\end{center}
\caption{The statistical average $\langle L_3\rangle$ of the shortest triangle side
$L_3$. We plot this quantity for a series of triangle mass bins 
and we divide $\langle L_3\rangle$ by the factor $\sqrt{M}$, where $M$ is the
mean mass of each bin.}
\label{figl3-2d}
\end{figure}

As discussed in the previous section an important ingredient for the derivation of the 
Lagrangian propagator is the triangle shape distribution function. In particular the 
derivation of Eq.(\ref{Rkappa-high-k-d-2-equil}) assumes that the triangles that build
the Lagrangian-space tessellation of the matter distribution 
\citep{2010PhRvE..82a6311B} are scale-invariant in the small-mass limit.
In other words, it neglects any dependence of their shape probability
distribution on scale since it assumes that typical lengths scale as the
square-root $\ell$ of the triangle area (in the limit of small triangles).
This assumption is violated if the shape distribution does not converge
to a finite limit, for instance if smaller triangles are increasingly ``squeezed''
as in Eq.(\ref{Rkappa-high-k-d-2-squeezed}).

To investigate this point, we consider in this section the dependence on scale
of the triangle geometry.
Thus, for each realization and at each output time, we label the three sides 
of each triangle of the Lagrangian-space tessellation according to
\beq
\ell_1 \geq \ell_2 \geq \ell_3 ,
\label{lside-def}
\eeq
that is, $\ell_1$ is the longest side and $\ell_3$ the shortest side.
Then, we plot in Figs.~\ref{figl1-2d} and \ref{figl3-2d} the averages of the ratios
$L_1/\sqrt{M}=\ell_1/\sqrt{\cal A}$ and $L_3/\sqrt{M}=\ell_3/\sqrt{\cal A}$,
where ${\cal A}=\ell^2$ is the triangle area.
To do so, we first bin the dimensionless triangle area $M$ (which is also the
dimensionless mass) over a finite number of bins. Then, for each realization we
compute the dimensionless lengths $L_1$ and $L_3$ and mass $M$ of all
triangles found in the Lagrangian-space tessellation at a given time, we
count all triangles that fall within a given mass bin and we also store their
lengths $L_1$ and $L_3$. Repeating this operation over several output times
(taking advantage of the self-similarity (\ref{selfsimilar})) and over many
initial conditions, characterized by the Gaussian statistics described in
Sect.~\ref{Initial-conditions}, we obtain the means $\langle L_1\rangle$ and
$\langle L_3\rangle$ within each mass bin, whence the ratios
$\langle L_1\rangle/\sqrt{M}$ and $\langle L_3\rangle/\sqrt{M}$ 
plotted in Figs.~\ref{figl1-2d} and \ref{figl3-2d}.
 
If the triangle distribution is scale-invariant in the small-mass limit (i.e. for
small area) these ratios must go to finite and nonzero constants at low $M$.
Figures \ref{figl1-2d} and \ref{figl3-2d} suggest that this is not the case, and
that $\langle L_1\rangle/\sqrt{M}$ keeps increasing while
$\langle L_3\rangle/\sqrt{M}$ keeps decreasing
at low mass. This means that triangles become increasingly ``squeezed'',
that is, $\ell_1\simeq \ell_2$ and $\ell_3\ll \ell_1$.
This behavior is consistent with a visual inspection of Fig.~6 in 
\cite{2010PhRvE..82a6311B}.
It also agrees with the analytical result obtained in \cite{1990MNRAS.246...10D}
when the velocity potential $\psi_0(\vq)$ is a random Poisson
point potential. However, this corresponds to very different initial conditions
than the ones studied in this paper, since then $\psi_0(\vq)$ is drawn from
a finite probability distribution at each point and there are no correlations
between different points. In contrast, in our case there are significant large-scale
correlations, which are characterized by the exponent $n$. As discussed in 
Sect.~\ref{Lagrangian-2d}, this leads to strong correlations in the Lagrangian-space
tessellation itself, with a combination of ``equilateral'' and ``squeezed'' triangles
that join summits that are in the same or two well-separated groups.

We can see in Figs.~\ref{figl1-2d} and \ref{figl3-2d} that 
the scale-dependence of the mean triangle shape is however quite weak and only logarithmic
(or possibly a power law with a small exponent), since the ratio
$\langle L_1\rangle/\sqrt{M}$ typically grows by a factor 5 and 
$\langle L_3\rangle/\sqrt{M}$ decreases by
a factor 2 when $M$ decreases by 4 orders of magnitude.
This would validate the scaling prediction (\ref{Rkappa-high-k-d-2-equil}), up to logarithmic
predictions. However, the deviations seen in Fig.~\ref{figdlRkappadlk-2d} suggest that
the mean values $\lag L_1\rag$ and $\lag L_3\rag$ are not sufficient to derive the
precise behavior of $\tR_{\kappa}$, which is quite sensitive to the relative fractions
and ``squeezed'' and ``equilateral'' configurations, and of the intermediate shapes,
since different exponents for $L_3$ as a function of $M$ give rise to different exponents
for the high-$K$ tail, in-between the two limiting cases 
(\ref{Rkappa-high-k-d-2-equil}) and (\ref{Rkappa-high-k-d-2-squeezed}). 
Nevertheless, the slightly steeper dependence on mass of the ratios
$\langle L_1\rangle/\sqrt{M}$ and $\langle L_3\rangle/\sqrt{M}$ observed for
larger $n$ agrees with Fig.~\ref{figdlRkappadlk-2d}, where the asymptotic slope
of the propagator appears to move farther from the ``equilateral'' prediction
for larger $n$.

Here we should note that these results differ from the properties associated with
the Delaunay triangulation of a Poisson point process, which are recalled in
App.~\ref{Delaunay}.
This is not surprising, since the Lagrangian-space tesselation is
not a Delaunay triangulation and the supporting points are non-Poisson distributed.
Thus, this discrepancy is another signature of the non-negligible correlations
that are the results of the power-law initial conditions (\ref{PdeltaL}), which
provide important large-scale correlations.

\section{Discussion and conclusion}
\label{Conclusion}

In this paper we have explored the concept of propagators for the geometrical
adhesion model, paying special attention to the Lagrangian propagators. As recalled
in the introduction propagators are defined in general as response functions of the final 
density field, or velocity field, to an infinitesimal change of the initial conditions. For 
Gaussian initial conditions these propagators can also be expressed in terms of unequal-time correlation 
functions. Following \cite{2007A&A...476...31V} we show in the first Section that this identity is quite general, 
more general than what was initially thought in \cite{2006PhRvD..73f3520C}, and can be derived 
beyond perturbation theory calculations.
%
%
%

The main result of this paper is the relation between the Lagrangian propagators
and the halo mass function that we uncovered in the context of the GAM.
The key point with which we established this connection is that, as soon as all the 
particles are gathered in halos, the displacement field in Lagrangian coordinates 
describes the formation of a partition where each cell correponds to a single halo.
Then, the convex hull construction that underlies the GAM explicitly shows that an 
infinitesimal variation of the initial conditions does not induce a change in this
Lagrangian partition, or in the mass of each halo, but only modifies the halo final 
positions. 
This can be understood by noticing that propagators are defined as ``linear'' response
functions, in the sense that they describe the sensitivity of the system to perturbations
up to linear order (but include the full nonlinear background dynamics). Although we did not do 
it explicitly it is clear that they can also be generalized to higher orders by expanding the response over powers of the
perturbation. It is also clear that this result is not specific to this model. It should be valid
as soon as the dynamics results in the partition of the Lagrangian space into cells
associated with distinct halos.

%

With the key property described above, the functional relation (\ref{tRkappa-W}) between the
halo mass function and the propagator  takes the form,
\begin{equation}
\tR_{\kappa}(k) \sim \int_{0}^{\infty}\dd m\, n(m)\,m\,\tW(k;m) .
\end{equation}
This provides an explicit link between the low-mass tail of the halo mass function
and the high-$k$ tail of the propagator. Moreover, the dimensionless kernel $\tW(k;m)$
contains some further information on the low-mass Lagrangian cells, more precisely
it describes the scalings with mass of their typical lengths (in dimension greater than
one there may be more than one relevant scale, as cells may become more or less
``squeezed'' or ``flattened''). This relationship provides a well-defined range of possible exponents for the 
asymptotic behavior of the Lagrangian propagator, if we know the low-mass tail of the
halo mass function, and a definite prediction if we also know the shape of low-mass 
cells.

In a more general context, such as the 3D gravitational dynamics, we cannot
derive an explicit relationship of this form. However, it is natural to expect
again a strong link between the Lagrangian propagators and the halo mass function.
This must be contrasted with the Eulerian propagators, where in both dynamics the main
process at play for these initial conditions is a ``sweeping effect'' due to
long-wavelenght modes of the velocity field.
Thus, our results strongly suggest that Lagrangian propagators are generally much better 
probes of the density field, and in particular of the halo mass function. 
Moreover, a new result that we have uncovered in this paper and that appears in
dimensions greater than one is the sensitivity to the shape of the underlying
Lagrangian-space tessellation, through the kernel $\tW(k;m)$, which should also remain
valid for more general cases.

It would be interesting to use the deeper understanding brought by such studies
to improve or build quantitative tools. 
With respect to the Lagrangian propagators two avenues naturally appear.
One could first use Lagrangian response functions to probe some properties of the
system. However, in a cosmological context where one cannot perform actual
experiments (except for numerical simulations) this may not be very practical.
A second route would be to use our understanding to build more efficient analytical
schemes, such as new resummation methods of Lagrangian perturbation theory.

\appendix

\section{Angle distribution in Delaunay triangulation from a Poisson point process}
\label{Delaunay}

We consider a set of points with a Poisson distribution in a $d=2$ space and the Delaunay 
triangulation associated with those points. It is then possible to derive the joint probability 
distribution  function of the three angles, $\alpha_{1}$, $\alpha_{2}$ and $\alpha_{3}=\pi-\alpha_{1}-\alpha_{2}$, of the triangles.
It is given by \cite{Miles197085,1987A&A...184...16I},
\begin{eqnarray}
P_{\rm tri.}(\alpha_{1},\alpha_{2})\dd\alpha_{1}\dd\alpha_{2}&=&\nonumber\\
&&\hspace{-3cm}
\frac{8}{3\pi}\sin(\alpha_{1})\sin(\alpha_{2})\sin(\alpha_{1}+\alpha_{2})\dd\alpha_{1}\dd\alpha_{2}
\end{eqnarray}
with
\begin{equation}
0<\alpha_{1}<\pi,\ 0<\alpha_{2}<\pi,\ 0<\alpha_{1}+\alpha_{2}<\pi.
\end{equation}
This distribution entirely characterizes the shape distribution of the triangles. In particular one can derive
the average length of each side for a given area. Each side length is given by
\begin{equation}
\frac{s_{3}^2}{\mA}=\frac{2\sin\alpha_{3}}{\sin\alpha_{1}\sin\alpha_{2}}
\end{equation}
where  $\mA$ is the surface area of the triangle and where the 2 other side lengths are obtained by circular 
permutations of the indices. The resulting average value of the longest and shortest
sides are then respectively,
\begin{equation}
\frac{\langle L_{1}\rangle}{\sqrt{\mA}}= 2.39298
\end{equation}
and
\begin{equation}
\frac{\langle L_{3}\rangle}{\sqrt{\mA}}= 1.12187 .
\end{equation}
These results are to be compared to those obtained in our construction.

\section{Real-space Eulerian propagators}
\label{real-space}

As seen in Eq.(\ref{Rpsi-pxu}) the real-space response function $\pR_{\psi}(\vx,t;\vq_0)$
of the velocity potential is given by the one-point velocity probability distribution function.
In terms of the reduced variables  (\ref{QXU}), using the statistical homogeneity of the system,
this reads as
\beq
\pR_{\psi}(\vX) = P(\vU) , \;\;\; \mbox{with} \;\;\; \vU=\vX ,
\label{Rpsi-PU-d}
\eeq
which holds in any dimension.
We show in this appendix our results in 1D and 2D for this propagator, or velocity distribution,
which is also the inverse Fourier transform of the propagators plotted in 
Figs.~\ref{figRpsi-1d} and \ref{figGdeltaGpsi-2d}.
To measure this real-space quantity we do not use the functional derivative (\ref{Rpsi-def}).
Instead, using Eq.(\ref{Rpsi-PU-d}) we simply measure the velocity distribution.

We can note that the identity (\ref{Rpsi-PU-d}) implies that the real-space propagator
$\pR_{\psi}$ is always positive. This is not the case for its Fourier-space counterpart
$\tR_{\psi}$. For instance, it was proved in \cite{2010PhRvD..81d3516B}
that in 1D for $n=0$ the far tail of $\tR_{\psi}(K)$ shows fast oscillations that are
exponentially damped (a first oscillation can be distinguished on the dot-dashed curve in
Fig.~\ref{figRpsi-1d}).

As in Sect.~\ref{EulerianPropagators} we only consider the initial conditions
$n=0.5, 0$, and $-0.5$, since as explained in Sect.~\ref{Eulerian-propagators}
these Eulerian propagators and one-point velocity distributions do not exist for
$n<-1$.

\subsection{1D case}
\label{real-space-1D}

\begin{figure}
\begin{center}
\epsfxsize=8.5 cm \epsfysize=6 cm {\epsfbox{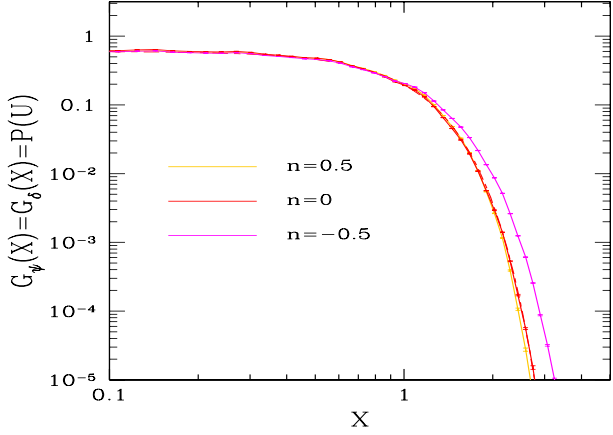}}
\end{center}
\caption{The real-space response function $\pR_{\psi}(X)=\pR_{\delta}(X)$
from 1D numerical simulations (solid line). We show the cases $n=0.5, 0$, and 
$-0.5$, as a function of the reduced distance $X$. For $n=0$ we also
plot the exact analytical result (dot-dashed line) \cite{2010PhRvD..81d3516B}.}
\label{figRpsix-1d}
\end{figure}

\begin{figure}
\begin{center}
\epsfxsize=8.5 cm \epsfysize=6 cm {\epsfbox{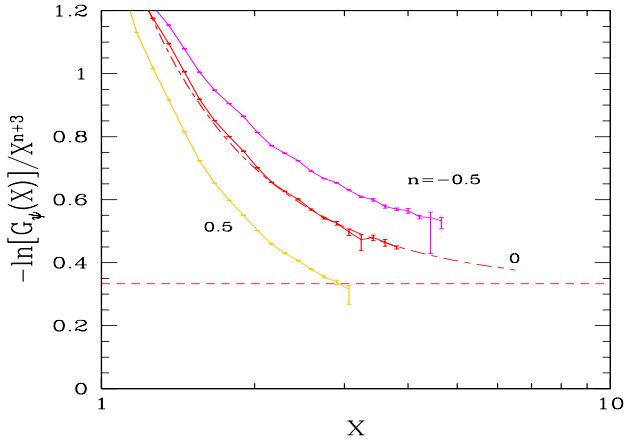}}
\end{center}
\caption{The ratio $-\ln [\pR_{\psi}(X)]/X^{(n+3)}$ from 1D numerical simulations
(solid line). For $n=0$ we also plot the exact analytical result (dot-dashed line)
and its asymptotic limit (dashed line) \cite{2010PhRvD..81d3516B}.}
\label{figlnRpsi_lnx-1d}
\end{figure}

We show in Fig.~\ref{figRpsix-1d} the real-space response
function $\pR_{\psi}(X)=\pR_{\delta}(X)$.
As in Sect.~\ref{One-dimensionalEulerian}, we can check that for $n=0$ our numerical 
result agrees very well with the exact result (which cannot even be distinguished in
this figure), given in 
\cite{2000JFM...417..323F,2009JSP...137..729V,2010PhRvD..81d3516B}.
The response functions obtained for $n=0.5$ and $-0.5$ show the same behavior,
with a sharp large-distance falloff, but contrary to its Fourier transform plotted in
Fig.~\ref{figRpsi-1d} we can clearly see the dependence on $n$
(especially for $n=-0.5$).

To clearly see this large-distance falloff, we plot in Fig.~\ref{figlnRpsi_lnx-1d} the
ratio $-\ln [\pR_{\psi}(X)]/X^{(n+3)}$, which must go to a constant at large $X$
from the analytical result (\ref{REul_asymp}). 
As in Fig.~\ref{figRpsix-1d}, we can check that for $n=0$ our numerical result
follows the exact result from
\cite{2000JFM...417..323F,2009JSP...137..729V,2010PhRvD..81d3516B}.
The curves obtained for $n=0.5$ and $-0.5$ display a similar behavior, but as shown
by the explicit case $n=0$ the curves have not converged to their asymptotic limit yet.
Therefore, we cannot precisely measure from the simulations the exponent
of the large-distance falloff, but they are consistent with the analytical result
(\ref{REul_asymp}). 
The values reached in Fig.~\ref{figlnRpsi_lnx-1d} correspond to $\pR_{\psi}(X)$ and
$P(U)$ below $10^{-10}$ for the case $n=0$, which means that the convergence
of the ratio $-\ln [\pR_{\psi}(X)]/X^{(n+3)}$ to its asymptotic limit is rather slow.

\subsection{2D case}
\label{real-space-2D}

\begin{figure}
\begin{center}
\epsfxsize=8.5 cm \epsfysize=6 cm {\epsfbox{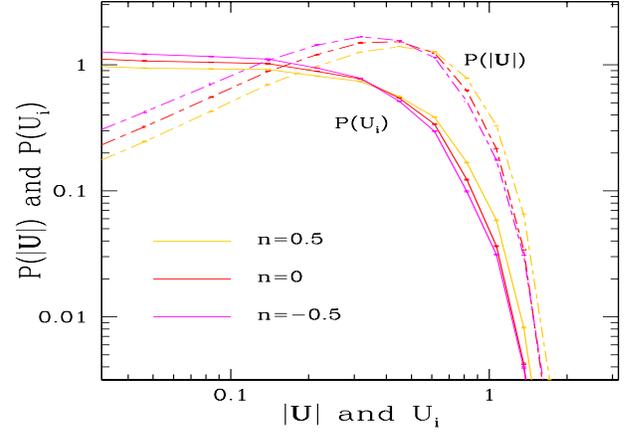}}
\end{center}
\caption{The real-space response function $\pR_{\psi}(\vX)=P(\vU)$, with
$\vX=\vU$. We show the probability distribution of each velocity component, $P(U_i)$
(solid line), and the probability distribution of the total amplitude,
$P(|\vU|)$ (dashed line).}
\label{figRpsix-2d}
\end{figure}

We show in Fig.~\ref{figRpsix-2d} the real-space response function $\pR_{\psi}(\vX)$,
or more precisely the velocity probability distribution $P(\vU)$, using the
identity (\ref{Rpsi-PU-d}). This is the inverse Fourier transform of the response
function $\tR_{\psi}(K)$ shown in Fig.~\ref{figGdeltaGpsi-2d}. 
As in the 1D case, we obtain
a sharp exponential-like cutoff at large distance (i.e. at large velocity in terms
of $P(\vU)$), which is consistent with Eq.(\ref{Rpsi-d-tail}).

\section{Cross-correlations}
\label{Cross-correlations}

\begin{figure}
\begin{center}
\epsfxsize=8.5 cm \epsfysize=6 cm {\epsfbox{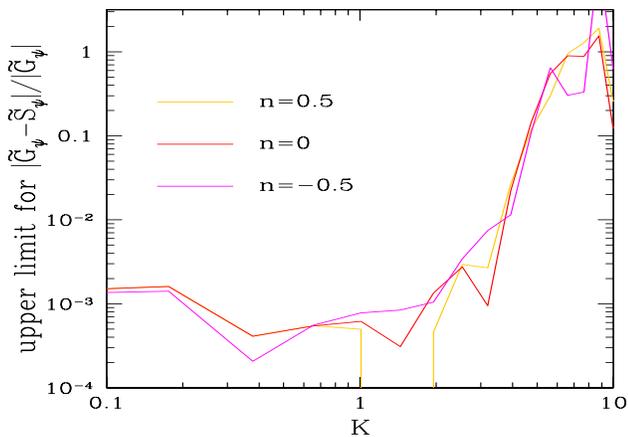}}
\end{center}
\caption{The upper limit for the relative difference
$|\tR_{\psi}-\tS_{\psi}|/|\tR_{\psi}|$, from 1D numerical simulations.
We show the cases $n=0.5, 0$, and $-0.5$, as in Fig.~\ref{figRpsi-1d}.}
\label{figdlRGpsi-1d}
\end{figure}

\begin{figure}
\begin{center}
\epsfxsize=8.5 cm \epsfysize=6 cm {\epsfbox{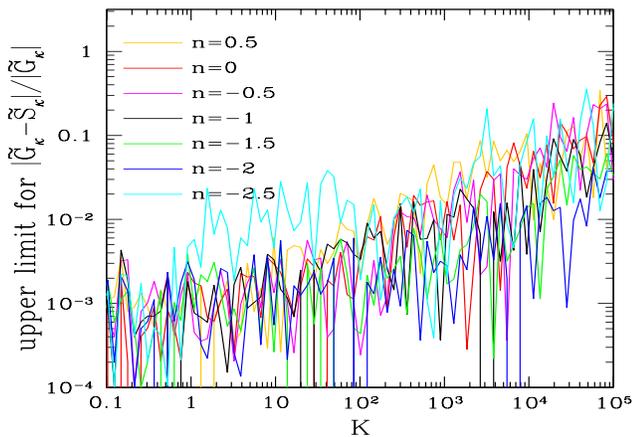}}
\end{center}
\caption{The upper limit for the relative difference 
$|\tR_{\kappa}-\tS_{\kappa}|/|\tR_{\kappa}|$, from 1D numerical simulations.}
\label{figdlRGkappa-1d}
\end{figure}

We show in Figs.~\ref{figdlRGpsi-1d} and \ref{figdlRGkappa-1d} the relative
difference between the response functions $\tR(K)$, computed through
functional derivatives as in Eq.(\ref{tR-def}), and the same quantities computed
through cross-correlations as in Eq.(\ref{tP-def}), which we denote as
$\tS(K)$ to distinguish between both procedures. 
We show our results for the 1D Eulerian propagator $\tR_{\psi}$ 
in Fig.~\ref{figdlRGpsi-1d}, and for the 1D Lagrangian propagator $\tR_{\kappa}$
in Fig.~\ref{figdlRGkappa-1d}.
We plot the upper limit that we obtain for the relative difference
$|\tR-\tS|/|\tR|$, since our errobars are everywhere consistent with the
exact equality $\tR=\tS$. (Of course this upper limit weakens at high $K$
where the propagators are very small and numerical measures are more
difficult.)
Wo obtain similar results in 2D but the tests of this identity are somewhat weaker
because of the lower-quality statistics.

These figures show the validity of our numerical algorithms and provide an
estimate of the accuracy of our measures of the propagators.
They also confirm the validity of the identities (\ref{S-def}) and (\ref{R-funct-2})
in the nonperturbative highly nonlinear regime, in agreement with the derivation presented
in the introduction.

\bibliography{refBurgers}   

\begin{thebibliography}{48}
\expandafter\ifx\csname natexlab\endcsname\relax\def\natexlab#1{#1}\fi
\expandafter\ifx\csname bibnamefont\endcsname\relax
  \def\bibnamefont#1{#1}\fi
\expandafter\ifx\csname bibfnamefont\endcsname\relax
  \def\bibfnamefont#1{#1}\fi
\expandafter\ifx\csname citenamefont\endcsname\relax
  \def\citenamefont#1{#1}\fi
\expandafter\ifx\csname url\endcsname\relax
  \def\url#1{\texttt{#1}}\fi
\expandafter\ifx\csname urlprefix\endcsname\relax\def\urlprefix{URL }\fi
\providecommand{\bibinfo}[2]{#2}
\providecommand{\eprint}[2][]{\url{#2}}

\bibitem[{\citenamefont{{Ostriker} and
  {Steinhardt}}(1995)}]{1995Natur.377..600O}
\bibinfo{author}{\bibfnamefont{J.~P.} \bibnamefont{{Ostriker}}}
  \bibnamefont{and} \bibinfo{author}{\bibfnamefont{P.~J.}
  \bibnamefont{{Steinhardt}}}, \bibinfo{journal}{\nat}
  \textbf{\bibinfo{volume}{377}}, \bibinfo{pages}{600} (\bibinfo{year}{1995}).

\bibitem[{\citenamefont{{Spergel} et~al.}(2003)\citenamefont{{Spergel},
  {Verde}, {Peiris}, {Komatsu}, {Nolta}, {Bennett}, {Halpern}, {Hinshaw},
  {Jarosik}, {Kogut} et~al.}}]{2003ApJS..148..175S}
\bibinfo{author}{\bibfnamefont{D.~N.} \bibnamefont{{Spergel}}},
  \bibinfo{author}{\bibfnamefont{L.}~\bibnamefont{{Verde}}},
  \bibinfo{author}{\bibfnamefont{H.~V.} \bibnamefont{{Peiris}}},
  \bibinfo{author}{\bibfnamefont{E.}~\bibnamefont{{Komatsu}}},
  \bibinfo{author}{\bibfnamefont{M.~R.} \bibnamefont{{Nolta}}},
  \bibinfo{author}{\bibfnamefont{C.~L.} \bibnamefont{{Bennett}}},
  \bibinfo{author}{\bibfnamefont{M.}~\bibnamefont{{Halpern}}},
  \bibinfo{author}{\bibfnamefont{G.}~\bibnamefont{{Hinshaw}}},
  \bibinfo{author}{\bibfnamefont{N.}~\bibnamefont{{Jarosik}}},
  \bibinfo{author}{\bibfnamefont{A.}~\bibnamefont{{Kogut}}},
  \bibnamefont{et~al.}, \bibinfo{journal}{\apjs}
  \textbf{\bibinfo{volume}{148}}, \bibinfo{pages}{175} (\bibinfo{year}{2003}),
  \eprint{arXiv:astro-ph/0302209}.

\bibitem[{\citenamefont{{Bernardeau} et~al.}(2002)\citenamefont{{Bernardeau},
  {Colombi}, {Gazta{\~n}aga}, and {Scoccimarro}}}]{2002PhR...367....1B}
\bibinfo{author}{\bibfnamefont{F.}~\bibnamefont{{Bernardeau}}},
  \bibinfo{author}{\bibfnamefont{S.}~\bibnamefont{{Colombi}}},
  \bibinfo{author}{\bibfnamefont{E.}~\bibnamefont{{Gazta{\~n}aga}}},
  \bibnamefont{and}
  \bibinfo{author}{\bibfnamefont{R.}~\bibnamefont{{Scoccimarro}}},
  \bibinfo{journal}{\physrep} \textbf{\bibinfo{volume}{367}},
  \bibinfo{pages}{1} (\bibinfo{year}{2002}), \eprint{arXiv:astro-ph/0112551}.

\bibitem[{\citenamefont{{Cooray} and {Sheth}}(2002)}]{2002PhR...372....1C}
\bibinfo{author}{\bibfnamefont{A.}~\bibnamefont{{Cooray}}} \bibnamefont{and}
  \bibinfo{author}{\bibfnamefont{R.}~\bibnamefont{{Sheth}}},
  \bibinfo{journal}{\physrep} \textbf{\bibinfo{volume}{372}},
  \bibinfo{pages}{1} (\bibinfo{year}{2002}), \eprint{astro-ph/0206508}.

\bibitem[{\citenamefont{{Crocce} and
  {Scoccimarro}}(2006{\natexlab{a}})}]{2006PhRvD..73f3519C}
\bibinfo{author}{\bibfnamefont{M.}~\bibnamefont{{Crocce}}} \bibnamefont{and}
  \bibinfo{author}{\bibfnamefont{R.}~\bibnamefont{{Scoccimarro}}},
  \bibinfo{journal}{\prd} \textbf{\bibinfo{volume}{73}},
  \bibinfo{pages}{063519} (\bibinfo{year}{2006}{\natexlab{a}}),
  \eprint{arXiv:astro-ph/0509418}.

\bibitem[{\citenamefont{{Crocce} and
  {Scoccimarro}}(2006{\natexlab{b}})}]{2006PhRvD..73f3520C}
\bibinfo{author}{\bibfnamefont{M.}~\bibnamefont{{Crocce}}} \bibnamefont{and}
  \bibinfo{author}{\bibfnamefont{R.}~\bibnamefont{{Scoccimarro}}},
  \bibinfo{journal}{\prd} \textbf{\bibinfo{volume}{73}},
  \bibinfo{pages}{063520} (\bibinfo{year}{2006}{\natexlab{b}}),
  \eprint{arXiv:astro-ph/0509419}.

\bibitem[{\citenamefont{{Bernardeau} et~al.}(2008)\citenamefont{{Bernardeau},
  {Crocce}, and {Scoccimarro}}}]{2008PhRvD..78j3521B}
\bibinfo{author}{\bibfnamefont{F.}~\bibnamefont{{Bernardeau}}},
  \bibinfo{author}{\bibfnamefont{M.}~\bibnamefont{{Crocce}}}, \bibnamefont{and}
  \bibinfo{author}{\bibfnamefont{R.}~\bibnamefont{{Scoccimarro}}},
  \bibinfo{journal}{\prd} \textbf{\bibinfo{volume}{78}},
  \bibinfo{pages}{103521} (\bibinfo{year}{2008}), \eprint{0806.2334}.

\bibitem[{\citenamefont{{Valageas}}(2007{\natexlab{a}})}]{2007A&A...465..725V}
\bibinfo{author}{\bibfnamefont{P.}~\bibnamefont{{Valageas}}},
  \bibinfo{journal}{\aap} \textbf{\bibinfo{volume}{465}}, \bibinfo{pages}{725}
  (\bibinfo{year}{2007}{\natexlab{a}}), \eprint{arXiv:astro-ph/0611849}.

\bibitem[{\citenamefont{{Valageas}}(2008)}]{2008A&A...484...79V}
\bibinfo{author}{\bibfnamefont{P.}~\bibnamefont{{Valageas}}},
  \bibinfo{journal}{\aap} \textbf{\bibinfo{volume}{484}}, \bibinfo{pages}{79}
  (\bibinfo{year}{2008}), \eprint{0711.3407}.

\bibitem[{\citenamefont{{Taruya} and {Hiramatsu}}(2008)}]{2008ApJ...674..617T}
\bibinfo{author}{\bibfnamefont{A.}~\bibnamefont{{Taruya}}} \bibnamefont{and}
  \bibinfo{author}{\bibfnamefont{T.}~\bibnamefont{{Hiramatsu}}},
  \bibinfo{journal}{\apj} \textbf{\bibinfo{volume}{674}}, \bibinfo{pages}{617}
  (\bibinfo{year}{2008}), \eprint{0708.1367}.

\bibitem[{\citenamefont{{Hiramatsu} and {Taruya}}(2009)}]{2009PhRvD..79j3526H}
\bibinfo{author}{\bibfnamefont{T.}~\bibnamefont{{Hiramatsu}}} \bibnamefont{and}
  \bibinfo{author}{\bibfnamefont{A.}~\bibnamefont{{Taruya}}},
  \bibinfo{journal}{\prd} \textbf{\bibinfo{volume}{79}},
  \bibinfo{pages}{103526} (\bibinfo{year}{2009}), \eprint{0902.3772}.

\bibitem[{\citenamefont{{Pietroni}}(2008)}]{2008JCAP...10..036P}
\bibinfo{author}{\bibfnamefont{M.}~\bibnamefont{{Pietroni}}},
  \bibinfo{journal}{\jcap} \textbf{\bibinfo{volume}{10}}, \bibinfo{pages}{36}
  (\bibinfo{year}{2008}), \eprint{0806.0971}.

\bibitem[{\citenamefont{{Okamura} et~al.}(2011)\citenamefont{{Okamura},
  {Taruya}, and {Matsubara}}}]{2011arXiv1105.1491O}
\bibinfo{author}{\bibfnamefont{T.}~\bibnamefont{{Okamura}}},
  \bibinfo{author}{\bibfnamefont{A.}~\bibnamefont{{Taruya}}}, \bibnamefont{and}
  \bibinfo{author}{\bibfnamefont{T.}~\bibnamefont{{Matsubara}}},
  \bibinfo{journal}{ArXiv e-prints}  (\bibinfo{year}{2011}),
  \eprint{1105.1491}.

\bibitem[{\citenamefont{{Pichon} and {Bernardeau}}(1999)}]{1999A&A...343..663P}
\bibinfo{author}{\bibfnamefont{C.}~\bibnamefont{{Pichon}}} \bibnamefont{and}
  \bibinfo{author}{\bibfnamefont{F.}~\bibnamefont{{Bernardeau}}},
  \bibinfo{journal}{\aap} \textbf{\bibinfo{volume}{343}}, \bibinfo{pages}{663}
  (\bibinfo{year}{1999}), \eprint{arXiv:astro-ph/9902142}.

\bibitem[{\citenamefont{{Valageas}}(2011)}]{2011A&A...526A..67V}
\bibinfo{author}{\bibfnamefont{P.}~\bibnamefont{{Valageas}}},
  \bibinfo{journal}{\aap} \textbf{\bibinfo{volume}{526}}, \bibinfo{pages}{A67+}
  (\bibinfo{year}{2011}), \eprint{1009.0106}.

\bibitem[{\citenamefont{{Gurbatov} et~al.}(1989)\citenamefont{{Gurbatov},
  {Saichev}, and {Shandarin}}}]{1989MNRAS.236..385G}
\bibinfo{author}{\bibfnamefont{S.~N.} \bibnamefont{{Gurbatov}}},
  \bibinfo{author}{\bibfnamefont{A.~I.} \bibnamefont{{Saichev}}},
  \bibnamefont{and} \bibinfo{author}{\bibfnamefont{S.~F.}
  \bibnamefont{{Shandarin}}}, \bibinfo{journal}{\mnras}
  \textbf{\bibinfo{volume}{236}}, \bibinfo{pages}{385} (\bibinfo{year}{1989}).

\bibitem[{\citenamefont{{Vergassola} et~al.}(1994)\citenamefont{{Vergassola},
  {Dubrulle}, {Frisch}, and {Noullez}}}]{1994A&A...289..325V}
\bibinfo{author}{\bibfnamefont{M.}~\bibnamefont{{Vergassola}}},
  \bibinfo{author}{\bibfnamefont{B.}~\bibnamefont{{Dubrulle}}},
  \bibinfo{author}{\bibfnamefont{U.}~\bibnamefont{{Frisch}}}, \bibnamefont{and}
  \bibinfo{author}{\bibfnamefont{A.}~\bibnamefont{{Noullez}}},
  \bibinfo{journal}{\aap} \textbf{\bibinfo{volume}{289}}, \bibinfo{pages}{325}
  (\bibinfo{year}{1994}).

\bibitem[{\citenamefont{Burgers}(1974)}]{0302.60048}
\bibinfo{author}{\bibfnamefont{J.}~\bibnamefont{Burgers}},
  \emph{\bibinfo{title}{{The nonlinear diffusion equation. Asymptotic solutions
  and statistical problems.}}} (\bibinfo{publisher}{{Dordrecht - Boston: D.
  Reidel Publishing Company. X, 173 p.}}, \bibinfo{year}{1974}).

\bibitem[{\citenamefont{{Zel'Dovich}}(1970)}]{1970A&A.....5...84Z}
\bibinfo{author}{\bibfnamefont{Y.~B.} \bibnamefont{{Zel'Dovich}}},
  \bibinfo{journal}{\aap} \textbf{\bibinfo{volume}{5}}, \bibinfo{pages}{84}
  (\bibinfo{year}{1970}).

\bibitem[{\citenamefont{{Bernardeau} and
  {Valageas}}(2010{\natexlab{a}})}]{2010PhRvE..82a6311B}
\bibinfo{author}{\bibfnamefont{F.}~\bibnamefont{{Bernardeau}}}
  \bibnamefont{and}
  \bibinfo{author}{\bibfnamefont{P.}~\bibnamefont{{Valageas}}},
  \bibinfo{journal}{\pre} \textbf{\bibinfo{volume}{82}},
  \bibinfo{pages}{016311} (\bibinfo{year}{2010}{\natexlab{a}}),
  \eprint{0912.3603}.

\bibitem[{\citenamefont{{Valageas} and
  {Bernardeau}}(2011)}]{2011PhRvD..83d3508V}
\bibinfo{author}{\bibfnamefont{P.}~\bibnamefont{{Valageas}}} \bibnamefont{and}
  \bibinfo{author}{\bibfnamefont{F.}~\bibnamefont{{Bernardeau}}},
  \bibinfo{journal}{\prd} \textbf{\bibinfo{volume}{83}},
  \bibinfo{pages}{043508} (\bibinfo{year}{2011}), \eprint{1009.1974}.

\bibitem[{\citenamefont{{Valageas}}(2007{\natexlab{b}})}]{2007A&A...476...31V}
\bibinfo{author}{\bibfnamefont{P.}~\bibnamefont{{Valageas}}},
  \bibinfo{journal}{\aap} \textbf{\bibinfo{volume}{476}}, \bibinfo{pages}{31}
  (\bibinfo{year}{2007}{\natexlab{b}}), \eprint{0706.2593}.

\bibitem[{\citenamefont{{Bernardeau} and
  {Valageas}}(2010{\natexlab{b}})}]{2010PhRvD..81d3516B}
\bibinfo{author}{\bibfnamefont{F.}~\bibnamefont{{Bernardeau}}}
  \bibnamefont{and}
  \bibinfo{author}{\bibfnamefont{P.}~\bibnamefont{{Valageas}}},
  \bibinfo{journal}{\prd} \textbf{\bibinfo{volume}{81}},
  \bibinfo{pages}{043516} (\bibinfo{year}{2010}{\natexlab{b}}),
  \eprint{0912.0356}.

\bibitem[{\citenamefont{{Bernardeau} and
  {Valageas}}(2008)}]{2008PhRvD..78h3503B}
\bibinfo{author}{\bibfnamefont{F.}~\bibnamefont{{Bernardeau}}}
  \bibnamefont{and}
  \bibinfo{author}{\bibfnamefont{P.}~\bibnamefont{{Valageas}}},
  \bibinfo{journal}{\prd} \textbf{\bibinfo{volume}{78}},
  \bibinfo{pages}{083503} (\bibinfo{year}{2008}), \eprint{0805.0805}.

\bibitem[{\citenamefont{Hopf}(1950)}]{0039.10403}
\bibinfo{author}{\bibfnamefont{E.}~\bibnamefont{Hopf}},
  \bibinfo{journal}{Commun. Pure Appl. Math.} \textbf{\bibinfo{volume}{3}},
  \bibinfo{pages}{201} (\bibinfo{year}{1950}).

\bibitem[{\citenamefont{Cole}(1951)}]{0043.09902}
\bibinfo{author}{\bibfnamefont{J.~D.} \bibnamefont{Cole}}, \bibinfo{journal}{Q.
  Appl. Math.} \textbf{\bibinfo{volume}{9}}, \bibinfo{pages}{225}
  (\bibinfo{year}{1951}).

\bibitem[{\citenamefont{{Bec} and {Khanin}}(2007)}]{2007PhR...447....1B}
\bibinfo{author}{\bibfnamefont{J.}~\bibnamefont{{Bec}}} \bibnamefont{and}
  \bibinfo{author}{\bibfnamefont{K.}~\bibnamefont{{Khanin}}},
  \bibinfo{journal}{\physrep} \textbf{\bibinfo{volume}{447}},
  \bibinfo{pages}{1} (\bibinfo{year}{2007}), \eprint{0704.1611}.

\bibitem[{\citenamefont{Gurbatov et~al.}(1990)\citenamefont{Gurbatov, Malakhov,
  and Saichev}}]{0753.76004}
\bibinfo{author}{\bibfnamefont{S.}~\bibnamefont{Gurbatov}},
  \bibinfo{author}{\bibfnamefont{A.}~\bibnamefont{Malakhov}}, \bibnamefont{and}
  \bibinfo{author}{\bibfnamefont{A.}~\bibnamefont{Saichev}},
  \emph{\bibinfo{title}{{Nonlinear random waves in media without dispersion.
  (Nelinejnye sluchajnye volny v sredakh bez dispersii.)}}}
  (\bibinfo{publisher}{{Sovremennye Problemy Fiziki. 81. Moskva: Nauka. 216 p.
  }}, \bibinfo{year}{1990}).

\bibitem[{\citenamefont{Woyczy\'nski}(1998)}]{0919.60004}
\bibinfo{author}{\bibfnamefont{W.~A.} \bibnamefont{Woyczy\'nski}},
  \emph{\bibinfo{title}{{Burgers-KPZ turbulence. G\"ottingen lectures.}}}
  (\bibinfo{publisher}{{Lecture Notes in Mathematics. 1700. Berlin: Springer.
  xi, 318 p.}}, \bibinfo{year}{1998}).

\bibitem[{\citenamefont{{Gurbatov} et~al.}(1997)\citenamefont{{Gurbatov},
  {Simdyankin}, {Aurell}, {Frisch}, and {T{\'o}th}}}]{1997JFM...344..339G}
\bibinfo{author}{\bibfnamefont{S.~N.} \bibnamefont{{Gurbatov}}},
  \bibinfo{author}{\bibfnamefont{S.~I.} \bibnamefont{{Simdyankin}}},
  \bibinfo{author}{\bibfnamefont{E.}~\bibnamefont{{Aurell}}},
  \bibinfo{author}{\bibfnamefont{U.}~\bibnamefont{{Frisch}}}, \bibnamefont{and}
  \bibinfo{author}{\bibfnamefont{G.}~\bibnamefont{{T{\'o}th}}},
  \bibinfo{journal}{Journal of Fluid Mechanics} \textbf{\bibinfo{volume}{344}},
  \bibinfo{pages}{339} (\bibinfo{year}{1997}), \eprint{arXiv:physics/9709002}.

\bibitem[{\citenamefont{{Valageas}}(2009{\natexlab{a}})}]{2009PhRvE..80a6305V}
\bibinfo{author}{\bibfnamefont{P.}~\bibnamefont{{Valageas}}},
  \bibinfo{journal}{\pre} \textbf{\bibinfo{volume}{80}},
  \bibinfo{pages}{016305} (\bibinfo{year}{2009}{\natexlab{a}}),
  \eprint{0905.1910}.

\bibitem[{\citenamefont{{Bernardeau} et~al.}(2011)\citenamefont{{Bernardeau},
  {Van de Rijt}, and {Vernizzi}}}]{2011arXiv1109.3400B}
\bibinfo{author}{\bibfnamefont{F.}~\bibnamefont{{Bernardeau}}},
  \bibinfo{author}{\bibfnamefont{N.}~\bibnamefont{{Van de Rijt}}},
  \bibnamefont{and}
  \bibinfo{author}{\bibfnamefont{F.}~\bibnamefont{{Vernizzi}}},
  \bibinfo{journal}{ArXiv e-prints}  (\bibinfo{year}{2011}),
  \eprint{1109.3400}.

\bibitem[{\citenamefont{{Frachebourg} and
  {Martin}}(2000)}]{2000JFM...417..323F}
\bibinfo{author}{\bibfnamefont{L.}~\bibnamefont{{Frachebourg}}}
  \bibnamefont{and} \bibinfo{author}{\bibfnamefont{P.~A.}
  \bibnamefont{{Martin}}}, \bibinfo{journal}{Journal of Fluid Mechanics}
  \textbf{\bibinfo{volume}{417}}, \bibinfo{pages}{323} (\bibinfo{year}{2000}),
  \eprint{arXiv:cond-mat/9905056}.

\bibitem[{\citenamefont{{Valageas}}(2009{\natexlab{b}})}]{2009JSP...137..729V}
\bibinfo{author}{\bibfnamefont{P.}~\bibnamefont{{Valageas}}},
  \bibinfo{journal}{Journal of Statistical Physics}
  \textbf{\bibinfo{volume}{137}}, \bibinfo{pages}{729}
  (\bibinfo{year}{2009}{\natexlab{b}}), \eprint{0903.0956}.

\bibitem[{\citenamefont{Molchan}(1997)}]{0944.60073}
\bibinfo{author}{\bibfnamefont{G.}~\bibnamefont{Molchan}}, \bibinfo{journal}{J.
  Stat. Phys.} \textbf{\bibinfo{volume}{88}}, \bibinfo{pages}{1139}
  (\bibinfo{year}{1997}).

\bibitem[{\citenamefont{Ryan}(1998)}]{0897.60075}
\bibinfo{author}{\bibfnamefont{R.}~\bibnamefont{Ryan}},
  \bibinfo{journal}{Commun. Math. Phys.} \textbf{\bibinfo{volume}{191}},
  \bibinfo{pages}{71} (\bibinfo{year}{1998}).

\bibitem[{\citenamefont{She et~al.}(1992)\citenamefont{She, Aurell, and
  Frisch}}]{0755.60104}
\bibinfo{author}{\bibfnamefont{Z.-S.} \bibnamefont{She}},
  \bibinfo{author}{\bibfnamefont{E.}~\bibnamefont{Aurell}}, \bibnamefont{and}
  \bibinfo{author}{\bibfnamefont{U.}~\bibnamefont{Frisch}},
  \bibinfo{journal}{Commun. Math. Phys.} \textbf{\bibinfo{volume}{148}},
  \bibinfo{pages}{623} (\bibinfo{year}{1992}).

\bibitem[{\citenamefont{Noullez and Vergassola}(1994)}]{0823.76058}
\bibinfo{author}{\bibfnamefont{A.}~\bibnamefont{Noullez}} \bibnamefont{and}
  \bibinfo{author}{\bibfnamefont{M.}~\bibnamefont{Vergassola}},
  \bibinfo{journal}{J. Sci. Comput.} \textbf{\bibinfo{volume}{9}},
  \bibinfo{pages}{259} (\bibinfo{year}{1994}).

\bibitem[{\citenamefont{{Brenier} et~al.}(2003)\citenamefont{{Brenier},
  {Frisch}, {H{\'e}non}, {Loeper}, {Matarrese}, {Mohayaee}, and {Sobolevski{\u
  \i}}}}]{2003MNRAS.346..501B}
\bibinfo{author}{\bibfnamefont{Y.}~\bibnamefont{{Brenier}}},
  \bibinfo{author}{\bibfnamefont{U.}~\bibnamefont{{Frisch}}},
  \bibinfo{author}{\bibfnamefont{M.}~\bibnamefont{{H{\'e}non}}},
  \bibinfo{author}{\bibfnamefont{G.}~\bibnamefont{{Loeper}}},
  \bibinfo{author}{\bibfnamefont{S.}~\bibnamefont{{Matarrese}}},
  \bibinfo{author}{\bibfnamefont{R.}~\bibnamefont{{Mohayaee}}},
  \bibnamefont{and}
  \bibinfo{author}{\bibfnamefont{A.}~\bibnamefont{{Sobolevski{\u \i}}}},
  \bibinfo{journal}{\mnras} \textbf{\bibinfo{volume}{346}},
  \bibinfo{pages}{501} (\bibinfo{year}{2003}), \eprint{arXiv:astro-ph/0304214}.

\bibitem[{\citenamefont{Sina\u\i}(1992)}]{0755.60105}
\bibinfo{author}{\bibfnamefont{Y.}~\bibnamefont{Sina\u\i}},
  \bibinfo{journal}{Commun. Math. Phys.} \textbf{\bibinfo{volume}{148}},
  \bibinfo{pages}{601} (\bibinfo{year}{1992}).

\bibitem[{\citenamefont{Bertoin}(1998)}]{0917.60063}
\bibinfo{author}{\bibfnamefont{J.}~\bibnamefont{Bertoin}},
  \bibinfo{journal}{Commun. Math. Phys.} \textbf{\bibinfo{volume}{193}},
  \bibinfo{pages}{397} (\bibinfo{year}{1998}).

\bibitem[{\citenamefont{{Valageas}}(2009{\natexlab{c}})}]{2009JSP...134..589V}
\bibinfo{author}{\bibfnamefont{P.}~\bibnamefont{{Valageas}}},
  \bibinfo{journal}{Journal of Statistical Physics}
  \textbf{\bibinfo{volume}{134}}, \bibinfo{pages}{589}
  (\bibinfo{year}{2009}{\natexlab{c}}), \eprint{0810.4332}.

\bibitem[{\citenamefont{Avellaneda and E}(1995)}]{0844.35144}
\bibinfo{author}{\bibfnamefont{M.}~\bibnamefont{Avellaneda}} \bibnamefont{and}
  \bibinfo{author}{\bibfnamefont{W.}~\bibnamefont{E}},
  \bibinfo{journal}{Commun. Math. Phys.} \textbf{\bibinfo{volume}{172}},
  \bibinfo{pages}{13} (\bibinfo{year}{1995}).

\bibitem[{\citenamefont{Miles}(1970)}]{Miles197085}
\bibinfo{author}{\bibfnamefont{R.}~\bibnamefont{Miles}},
  \bibinfo{journal}{Mathematical Biosciences} \textbf{\bibinfo{volume}{6}},
  \bibinfo{pages}{85 } (\bibinfo{year}{1970}), ISSN \bibinfo{issn}{0025-5564},
  \urlprefix\url{http://www.sciencedirect.com/science/article/pii/002555647090%
0611}.

\bibitem[{\citenamefont{{Icke} and {van de
  Weygaert}}(1987)}]{1987A&A...184...16I}
\bibinfo{author}{\bibfnamefont{V.}~\bibnamefont{{Icke}}} \bibnamefont{and}
  \bibinfo{author}{\bibfnamefont{R.}~\bibnamefont{{van de Weygaert}}},
  \bibinfo{journal}{\aap} \textbf{\bibinfo{volume}{184}}, \bibinfo{pages}{16}
  (\bibinfo{year}{1987}).

\bibitem[{\citenamefont{{Kofman} et~al.}(1990)\citenamefont{{Kofman},
  {Pogosian}, and {Shandarin}}}]{1990MNRAS.242..200K}
\bibinfo{author}{\bibfnamefont{L.}~\bibnamefont{{Kofman}}},
  \bibinfo{author}{\bibfnamefont{D.}~\bibnamefont{{Pogosian}}},
  \bibnamefont{and}
  \bibinfo{author}{\bibfnamefont{S.}~\bibnamefont{{Shandarin}}},
  \bibinfo{journal}{Mont. Not. Roy. Astron. Soc.}
  \textbf{\bibinfo{volume}{242}}, \bibinfo{pages}{200} (\bibinfo{year}{1990}).

\bibitem[{\citenamefont{{Press} and {Schechter}}(1974)}]{1974ApJ...187..425P}
\bibinfo{author}{\bibfnamefont{W.~H.} \bibnamefont{{Press}}} \bibnamefont{and}
  \bibinfo{author}{\bibfnamefont{P.}~\bibnamefont{{Schechter}}},
  \bibinfo{journal}{\apj} \textbf{\bibinfo{volume}{187}}, \bibinfo{pages}{425}
  (\bibinfo{year}{1974}).

\bibitem[{\citenamefont{{Doroshkevich} and
  {Kotok}}(1990)}]{1990MNRAS.246...10D}
\bibinfo{author}{\bibfnamefont{A.~G.} \bibnamefont{{Doroshkevich}}}
  \bibnamefont{and} \bibinfo{author}{\bibfnamefont{T.~V.}
  \bibnamefont{{Kotok}}}, \bibinfo{journal}{\mnras}
  \textbf{\bibinfo{volume}{246}}, \bibinfo{pages}{10} (\bibinfo{year}{1990}).

\end{thebibliography}

\end{document}